# *GoodRegressor*: A Hierarchical Inductive Bias for Navigating High-Dimensional Compositional Space

*Seong-Hoon Jang*[*1]

[1] Institute for Materials Research (IMR), Tohoku University, Sendai, 980-8577

*Corresponding author: jang.seonghoon.b4@tohoku.ac.jp (S.-H. Jang)

ABSTRACT

Interpretable scientific machine learning often trades predictive performance for structural transparency. When physical targets arise from hierarchical and nonlinear descriptor entanglement, weakly interacting white-box models underfit, whereas highly expressive black-box models obscure physical insight. **Here I introduce *GoodRegressor*, a hierarchical depth-controlled symbolic regression framework that systematically assembles nonlinear descriptor interactions through lexicographically-ordered expansion.** Despite effective compositional search spaces approaching $\sim 10^{400}$ structures, disciplined depth control enables tractable and reproducible exploration under realistic computational constraints. Across oxygen-ion conductors, NASICONs, and superconducting oxides, as representative high-complexity testbeds, predictive performances match or exceed state-of-the-art black-box models, retaining explicit functional form. Moreover, interaction-depth evolution reveals system-dependent optimal windows, providing an empirical taxonomy of hierarchical complexity in scientific datasets. **These results establish hierarchical inductive bias with explicit depth control as a design principle for interpretable artificial intelligence in high-dimensional compositional spaces, and position interaction depth as a structural axis for diagnosing hierarchical complexity in scientific systems.**

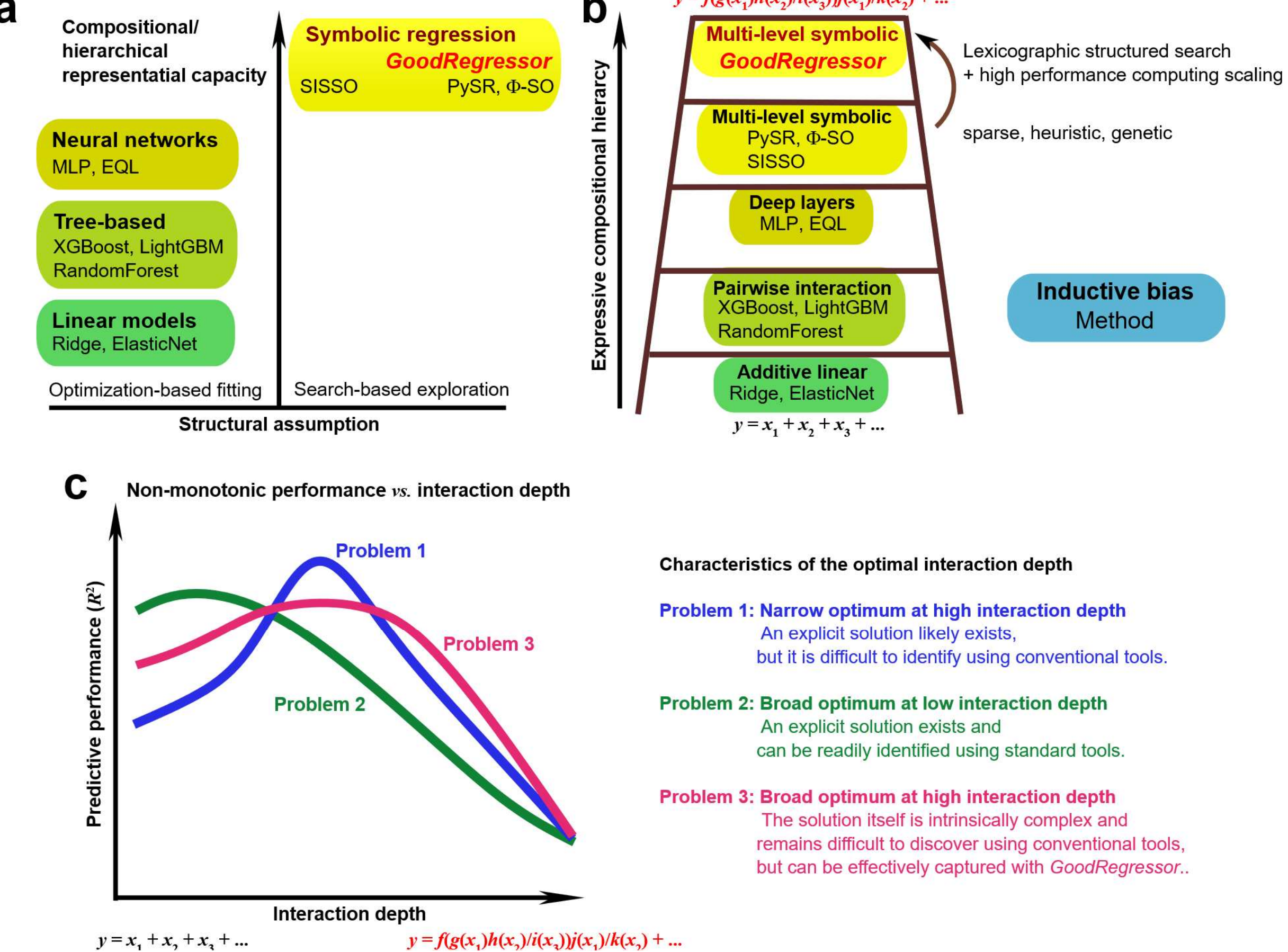

**Fig. 1** Inductive bias and hierarchical compositional structure in regression models. (a) Regression paradigms positioned by structural assumption (optimization-based fitting vs. search-based exploration) and compositional/hierarchical representational capacity. (b) Jacob's ladder illustrating increasing hierarchical symbolic expressivity, culminating in systematic multi-level construction. Each rung is annotated with representative methods, indicating the level of inductive bias they impose on how descriptors are structured to explain the data. (c) Non-monotonic performance as a function of interaction depth. Different systems exhibit distinct optimal regimes, ranging from shallow, easily accessible solutions to deep, intrinsically complex structures that are difficult to discover with conventional methods but can be captured with *GoodRegressor*.

Scientific artificial intelligence (AI) must reconcile interpretability with the hierarchical complexity of physical phenomena. Black-box models such as linear models[1, 2], neural networks,[3] random forests,[4] and gradient boosting[5, 6] often achieve strong predictive accuracy, yet their internal representations remain opaque and difficult to translate into physically meaningful insight. In contrast, white-box approaches provide explicit functional forms that allow analytical interpretation.[7-11] However, many of these approaches impose structural assumptions that are too restrictive to capture deeply entangled nonlinear interactions among descriptors.

Many scientific systems are governed by high-dimensional compositional interactions, where macroscopic properties emerge from complex combinations of underlying descriptors. Understanding and modeling such systems remains a central challenge in computational science. **Materials systems represent an extreme testbed for such compositional learning, as their target properties arise from descriptors that interact hierarchically and nonlinearly in highly entangled ways.** Additive or weakly interacting formulations often fail to capture essential coupling effects, whereas highly expressive models risk instability, overfitting, and loss of interpretability. Allowing descriptors to interact alleviates structural rigidity but introduces a fundamental combinatorial challenge. As hierarchical interaction depth increases, the number of possible descriptor compositions grows explosively, with effective search spaces reaching magnitudes on the order of $\sim 10^{400}$ candidate structures in realistic datasets, as demonstrated in the case studies below. **The central problem is therefore not merely the size of the combinatorial space, but how to traverse it under explicit structural bias within realistic computational constraints.**

To address this challenge, **I formalize hierarchical inductive bias with explicit depth control as a design principle for interpretable scientific AI.** Rather than searching expression space through unconstrained stochastic mutation or relying solely on parameter optimization within predefined templates, the proposed framework enforces structured, lexicographically ordered multi-level construction of nonlinear descriptor interactions. **Interaction depth serves as a controllable axis of compositional expressivity, functioning as a structural capacity axis that governs the symbolic hypothesis space, analogous to depth in neural network architectures, while enabling systematic exploration of increasingly entangled structures under analytical transparency and computational tractability.**

**Fig. 1a** situates representative regression paradigms within this conceptual landscape. The horizontal axis distinguishes optimization-based parameter fitting from structural search over candidate functional forms. Models such as linear models, tree-based models, and neural networks primarily optimize parameters within predefined structural templates. In contrast, symbolic regression explicitly searches over functional compositions themselves. The vertical axis represents compositional capacity, that is, the extent to which a model can assemble descriptors into hierarchical nonlinear structures. Linear models occupy the additive lower region. Tree-based methods relax strict additivity via piecewise partitioning but do not construct explicit algebraic compositions. Neural networks increase hierarchical capacity through layered nonlinear transformations, although the resulting representations remain implicit. Symbolic regression resides in the structural-search regime, explicitly assembling descriptor interactions.

Among symbolic regression frameworks, structural hypotheses are generated under distinct inductive biases. SISSO couples combinatorial descriptor construction with sparsity-driven optimization, occupying a middle ground between parameter fitting and structural search.[8] PySR and Φ-SO rely on stochastic evolutionary mutation and crossover to explore expression space.[9, 10] In contrast, I demonstrate that *GoodRegressor* employs hierarchical depth-controlled construction to assemble nonlinear interactions through structured, lexicographically ordered expansion, thereby enforcing explicit compositional control rather than relying primarily on stochastic variation. **In many symbolic regression frameworks, compositional depth is not treated as an explicit structural control axis, but arises implicitly from feature expansion or stochastic exploration. In contrast, *GoodRegressor* treats interaction depth as an explicit structural axis that governs hypothesis generation itself.**

Within this space, hierarchical depth-controlled symbolic construction embodies a distinct inductive bias. **Fig. 1b** conceptualizes compositional expansion as a "ladder" of expressivity, where each rung corresponds to increasing interaction depth. Importantly, ascending this ladder does not guarantee monotonic improvement in predictive performance. Instead, performance exhibits a non-monotonic, system-dependent optimum at intermediate interaction depth (**Fig. 1c**), reflecting an intrinsic trade-off between structural entanglement and generalization performance.

In the following sections, I demonstrate that interaction-depth evolution serves not only as a model-selection mechanism but also as a structural diagnostic of scientific datasets. Across oxygen-ion conductors,[12] NASICONs (Na-ion super ionic conductors) systems,[13] and superconducting oxides materials,[14] predictive performance remains comparable to state-of-the-art black-box models while preserving explicit symbolic form. **Here, materials systems are employed as representative high-complexity testbeds to probe this structural behavior.** More significantly, each system exhibits a characteristic interaction-depth signature, revealing how hierarchical complexity is organized in different high-dimensional compositional spaces. **This work therefore shifts symbolic regression from an algorithmic competition toward a question of inductive-bias design: how hierarchical compositional structure should be organized to render scientific complexity interpretable.**

## RESULTS

### Benchmark Performance with Conventional Machine Learning Approaches: a Case Study of Oxygen-ion Conductor Dataset

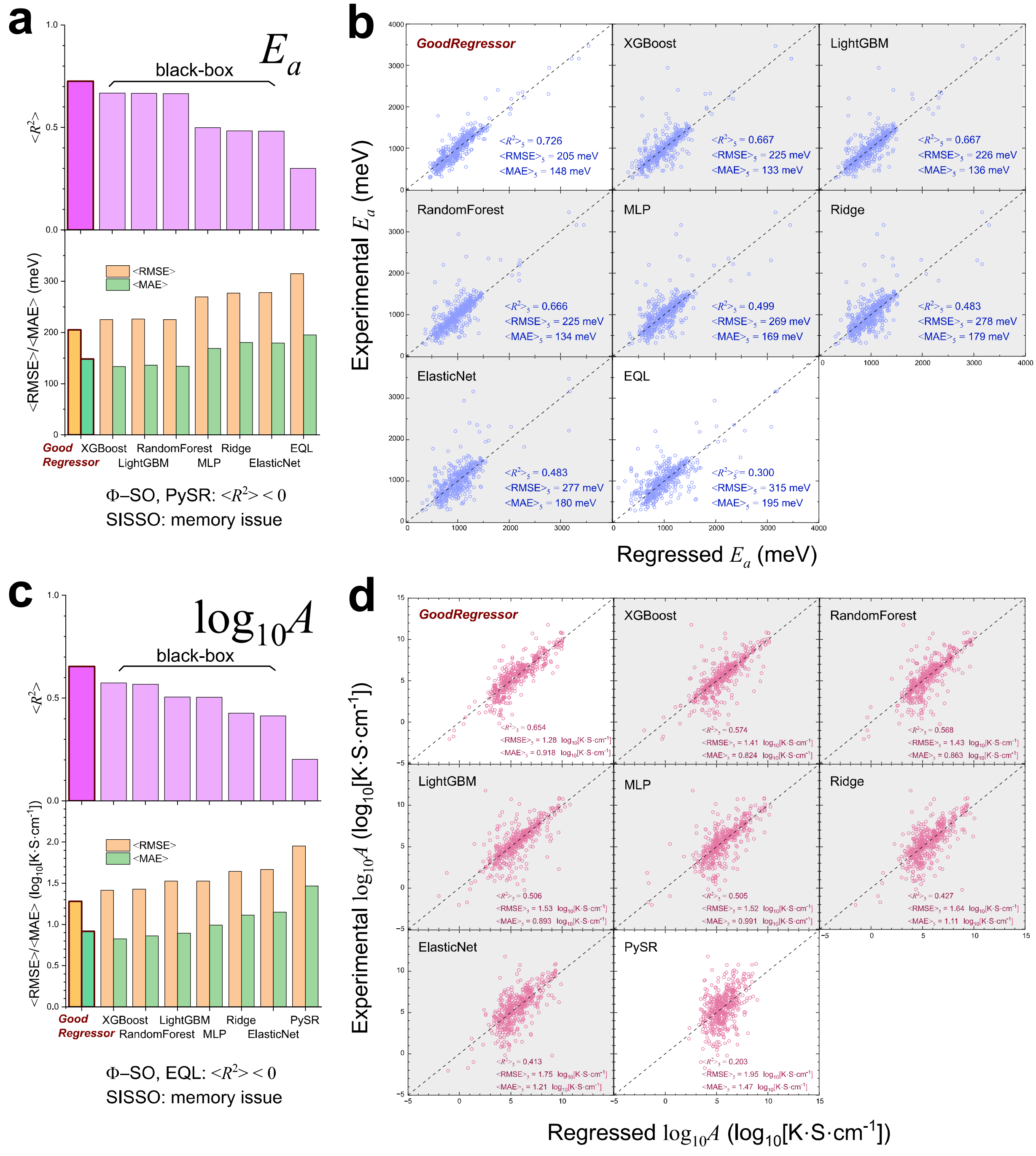

**Fig. 2** Benchmark performance of *GoodRegressor* and other machine learning models. Comparison of stacking-ensembled symbolic regression models with conventional machine learning approaches, Ridge,[1] ElasticNet,[2] MLP,[3] RandomForest,[4] XGBoost,[5] LightGBM,[6] EQL,[7] SISSO,[8] PySR,[9] and Φ-SO,[10] for predicting (a, b) activation energy ($E_a$) and (c, d) pre-exponential factor ($A$). (a) and (c) show averaged benchmark metrics across different test folds: coefficient of determination ($\langle R^2 \rangle$), root mean square error ($\langle$RMSE$\rangle$), and mean absolute error ($\langle$MAE$\rangle$). (b) and (d) present parity plots comparing experimental and predicted values for each model. In (b) and (d), the black-box models are illustrated as gray panels for visual distinction.

The technical details of the workflow are provided in the section **"Workflow"** in the **Supplementary Information.** Briefly, within the overall workflow, the regressor module independently generates $N_f$ symbolic models, $M_{f,i_f}$, by varying the random train-validation splits with an 8:2 ratio (bagging). Here, $i_f$ denotes the iteration index ($i_f = 1, \dots, N_f$), and each model is constructed using the optimized number ($n_t$) of active independent variables identified through the interaction-depth search, starting from the initial candidate feature pool $\mathbf{X_1}$. Subsequently, the designer module is employed as a post-processing tool for the symbolic regression results. Aggregating the ensemble of models up to iteration $i_f$ yields a consensus model, $\overline{M_{f,i_f}} = m_0 + \sum_{k_f=1}^{i_f} m_{k_f} M_{f,k_f}$ where the intercept $m_0$ and the coefficients $m_{k_f}$ are determined via least-squares fitting to the target dependent variable. In this study, it is given that $n_f = 10$.

Calculating $\overline{M_{f,10}}$, benchmark tests were conducted to compare their performance with other machine learning approaches for the same oxygen-ion conductor dataset (the number of data points $n_{\text{data}} = 483$ and the initial number of candidate independent variables $n(\mathbf{X_1}) = 358$; see **Supplementary Table 1**).[12] The interaction terms, according to the algorithm described in the section **"Workflow"** in the **Supplementary Information**, allows the model optimization among the combination number of $N_3^{\text{V}} = 1.44 \times 10^{457}$.

The comparison included black-box models, Ridge,[1] ElasticNet,[2] MLP,[3] RandomForest,[4] XGBoost,[5] and LightGBM,[6] as well as a white-box (symbolic regression) model, SISSO,[8] PySR,[9] and Φ-SO,[10] as illustrated in **Fig. 2**. Technical details are provided in the section **"Details of Conventional Machine Learning Approaches"** in the **Supplementary Information**. All models were trained and validated using 5-fold cross-validation, with hyperparameter optimization performed to enable fair and unbiased comparisons; the same folds were used across all the approaches. For instance, using *GoodRegressor*, in each iteration, the $n$-th 20 % fold of the dataset was held out during model construction, while the remaining folds ($k \neq n$) were used for training and validation. Within these remaining data, 10 independent random 8: 2 train-validation splits were generated to build one stacking-ensembled symbolic regression model. After model construction, the held-out $n$-th fold was used as the test set. This procedure yielded 50 individual symbolic models and 5 stacking-ensembled symbolic models in total, corresponding to overall data splits of 64 % training, 16 % validation, and 20 % testing.

In **Fig. 2a**, the benchmark results for the $E_a$ dataset are presented in terms of three metrics: the averaged coefficient of determination ($\langle R^2 \rangle_5$), the averaged root mean square error ($\langle \text{RMSE} \rangle_5$), and the averaged mean absolute error ($\langle \text{MAE} \rangle_5$) across the five validation folds. **The *GoodRegressor* model outperformed all other machine learning competitors.** Specifically, it achieved $\langle R^2 \rangle_5 = 0.726$, $\langle \text{RMSE} \rangle_5 = 205$ meV, and $\langle \text{MAE} \rangle_5 = 148$ meV. In contrast, the other models exhibited inferior performance ($\langle R^2 \rangle_5 \leq 0.667$, $\langle \mathrm{RMSE} \rangle_5 \geq 225$ meV, and $\langle \mathrm{MAE} \rangle_5 \geq 133$ meV). The performance ranking among these was as follows: XGBoost, LightGBM, RandomForest, MLP, Ridge, ElasticNet, and EQL. The parity plots in **Fig. 2b** further confirm that the *GoodRegressor* model provides excellent predictive accuracy. Meanwhile, EQL, the symbolic regression modeling baseline, failed to accurately reproduce the experimental values due to the astronomically large search space of possible symbolic expressions, which is a challenge effectively mitigated by *GoodRegressor*. For other symbolic baselines, it is worth noting that 60 training cycles (referred to as "epochs") in Φ-SO and PySR produced models with an $\langle R^2 \rangle_5$ value of only $\leq 0$. In addition, SISSO required more than 1.5 GB per core on a high-performance computing system (the same architecture employed for *GoodRegressor*) even when the "maximal feature complexity", defined as the number of operators per feature, was limited to 2 and the "S-expression" was adopted for the memory option. Consequently, SISSO was unable to handle the large feature space $n(\mathbf{X_1}) = 358$. For these reasons, the three unsuccessful cases are not shown in **Fig. 2b**.

**Fig. 2c** presents the benchmark results for the dataset of $A$. Again, the *GoodRegressor* model clearly outperforms the other machine learning methods. Specifically, it achieved $\langle R^2 \rangle_5 = 0.654$, $\langle \mathrm{RMSE} \rangle_5 = 1.28 \ \log_{10}[\mathrm{K} \cdot \mathrm{S} \cdot \mathrm{cm}^{-1}]$, and $\langle \mathrm{MAE} \rangle_5 = 0.918 \ \log_{10}[\mathrm{K} \cdot \mathrm{S} \cdot \mathrm{cm}^{-1}]$. The other models recorded lower performance ($\langle R^2 \rangle_5 \leq 0.574$, $\langle \mathrm{RMSE} \rangle_5 \geq 1.41 \ \log_{10}[\mathrm{K} \cdot \mathrm{S} \cdot \mathrm{cm}^{-1}]$, and $\langle \mathrm{MAE} \rangle_5 \geq 0.824 \ \log_{10}[\mathrm{K} \cdot \mathrm{S} \cdot \mathrm{cm}^{-1}]$). The performance ranking among these was as follows: XGBoost, RandomForest, LightGBM, MLP, Ridge, ElasticNet, and PySR. The benchmark results for Φ-SO, EQL, and SISSO are not presented for the same reasons outlined above. The parity plots in **Fig. 2d** further confirm that $\overline{M_{f,10}}$ provides excellent predictive accuracy as well.

## Stacking-ensembled Models $\overline{M_{f,\iota_f}}$ for $E_a$ and $A$

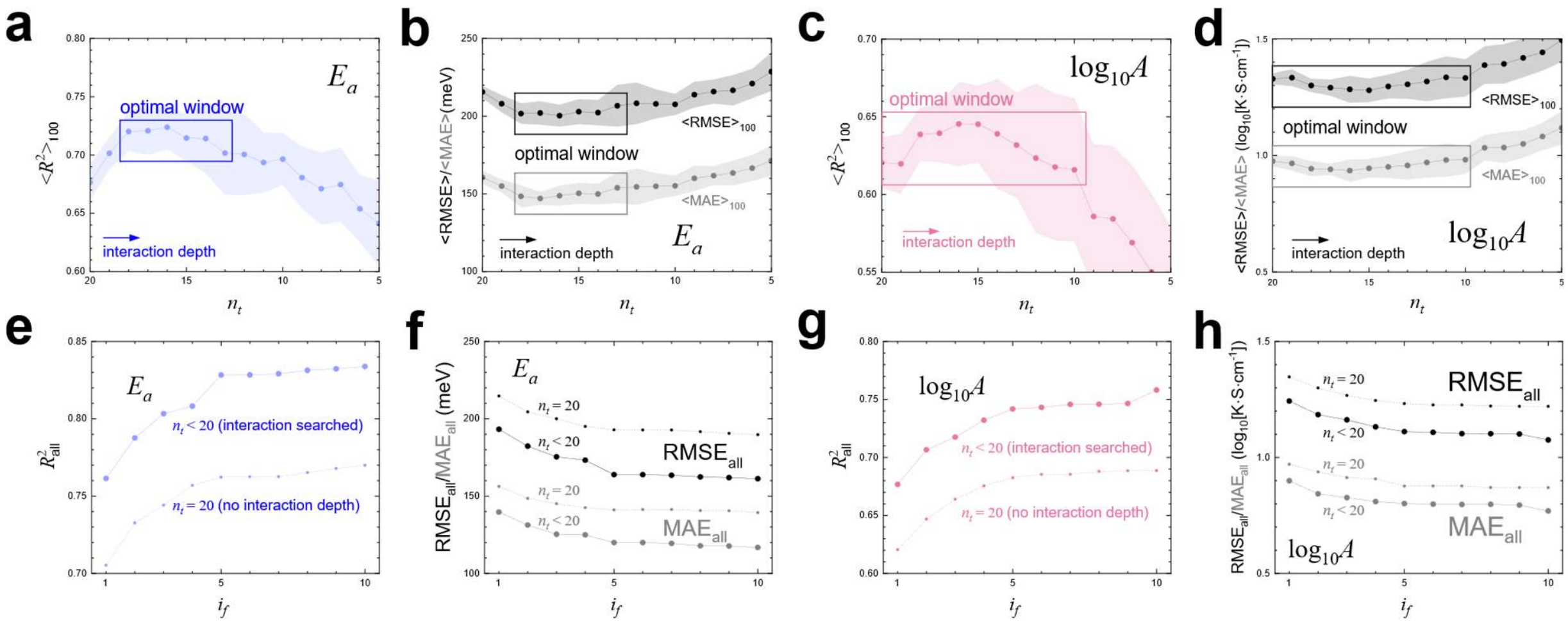

**Fig. 3** Interaction-depth evolution and stacking-ensemble performance for the oxygen-ion conductor case study. (a) Evolution of the cross-validated $\langle R^2 \rangle_{100}$ for $E_a$ with increasing hierarchical interaction depth, highlighting the optimal interaction window. The solid line represents the ensemble-averaged value, and the shaded region indicates the corresponding standard deviation across the 10 ensemble symbolic models. (b) Corresponding evolution of $\langle \mathrm{RMSE} \rangle_{100}$ and $\langle \mathrm{MAE} \rangle_{100}$ for $E_a$. (c) Evolution of $\langle R^2 \rangle_{100}$ for $\log_{10} A$. (d) Corresponding evolution of $\langle \mathrm{RMSE} \rangle_{100}$ and $\langle \mathrm{MAE} \rangle_{100}$ for $\log_{10} A$. (e-l) Evolution of the stacking-ensembled models $\overline{M_{f,\iota_f}}$ with increasing $i_f$ from 1 to 10. Shown are the saturation behaviors of (e) $R^2_{\mathrm{all}}$ and (f) $\mathrm{RMSE}_{\mathrm{all}}$ and $\mathrm{MAE}_{\mathrm{all}}$ for $E_a$, and those of (g) $R^2_{\mathrm{all}}$ and (h) $\mathrm{RMSE}_{\mathrm{all}}$ and $\mathrm{MAE}_{\mathrm{all}}$ for $\log_{10} A$. Dashed lines in panels (e–h) represent the corresponding stacking performance without hierarchical interaction construction (fixed $n_t = 20$), providing an ablation baseline for comparison.

Motivated by the strong predictive performance of *GoodRegressor* observed above, new stacking-ensembled models $\overline{M_{f,\iota_f}}$ were generated using the entire dataset with an 8: 2 train-validation split, without reserving a separate test set. As George E. P. Box famously noted, "all models are wrong"; reliance on a single model such as $M_{f,1}$ may overemphasize specific descriptors and distort interpretation of the underlying mechanisms. While, as examples, I leave the details of $M_{f,1}$ for $E_a$ and $A$ in the sections **Models $M_{f,1}$ for $E_a$ and $A$** in the **Supplementary Information**, **a stacking ensemble enables construction of a consensus model $\overline{M_{f,\iota_f}}$ that more faithfully captures the collective role of relevant descriptors.**

To begin, I analyze the optimization of the individual ensemble models that later serve as components of the stacking ensemble. **Figs. 3a-3d** illustrate the evolution of predictive performance as hierarchical interaction depth increases. As the effective interaction depth grows ($n_t = 20 \rightarrow 19 \rightarrow \cdots$), the cross-validated values of $\langle R^2 \rangle_{100}$ initially improve and reach the optima within the windows of $18 \geq n_t \geq 13$ for $E_a$ and $20 \geq n_t \geq 10$ and $\log_{10} A$. Beyond these ranges, $\langle R^2 \rangle_{100}$ decline (**Figs. 3a and 3c**), indicating that excessive compositional complexity deteriorates generalization performance. The corresponding trends in $\langle \mathrm{RMSE} \rangle_{100}$ and $\langle \mathrm{MAE} \rangle_{100}$ (**Figs. 3b and 3d**) mirror this behavior, reaching minima near the same interaction window and increasing thereafter. This non-monotonic evolution demonstrates that hierarchical expressivity alone does not guarantee improved predictability. Instead, there exists an optimal interaction window in which nonlinear descriptor entanglement enhances explanatory power without inducing noise-fitting. This empirical optimum corresponds to a specific rung of the compositional ladder introduced in **Fig. 1b**.

The overall evolution of $\overline{M_{f,i_f}}$, with increasing $i_f$ from 1 to $n_f = 10$, is illustrated in **Figs. 3e-3h**. As shown in **Fig. 3e**, for $E_a$, $R^2_{\mathrm{all}}$ increases sharply when $i_f$ increases from 1 to 2, and then rises more gradually thereafter. Between $i_f = 5$ and $10$, the improvement in $R^2_{\mathrm{all}}$ becomes negligible ($\Delta R^2_{\mathrm{all}} = 0.006$). Correspondingly, as illustrated in **Fig. 3f**, both $\mathrm{RMSE_{all}}$ and $\mathrm{MAE_{all}}$ decrease sharply from $i_f = 1$ to 2, followed by a gradual decline. Between $i_f = 5$ and 10, the reduction in $\mathrm{RMSE_{all}}$ ($\mathrm{MAE_{all}}$) is not significant, with $\Delta\mathrm{RMSE_{all}} = -2$ meV ($\Delta\mathrm{MAE_{all}} = -3$ meV). These results indicate that increasing the number of models beyond $i_f = 10$ offers no substantial benefit for $E_a$. **Importantly, the dashed-line baselines corresponding to models without hierarchical interaction construction ($\boldsymbol{n_t = 20}$) remain consistently inferior, demonstrating that the observed gains arise from structured interaction expansion rather than stacking alone.** Similarly, as shown in **Fig. 3g**, for $\log_{10} A$, $R^2_{\mathrm{all}}$ exhibits a sharp rise from $i_f = 1$ to 2, followed by a gradual increase. From $i_f = 5$ and $10$, $R^2_{\mathrm{all}}$ converges with only minor variation ($\Delta R^2_{\mathrm{all}} = 0.016$). As shown in **Fig. 3h**, $\mathrm{RMSE_{all}}$ and $\mathrm{MAE_{all}}$ also decrease rapidly up to $i_f = 2$, and then slowly stabilize. Between $i_f = 5$ and 10, the changes are insignificant: $\Delta\mathrm{RMSE_{all}} = -0.04 \log_{10}[\mathrm{K \cdot S \cdot cm^{-1}}]$ and $\Delta\mathrm{MAE_{all}} = -0.03 \log_{10}[\mathrm{K \cdot S \cdot cm^{-1}}]$. Again, the dashed-line baselines confirm that disabling hierarchical interaction depth results in consistently weaker performance.

One of the principal advantages of symbolic regression models is their capacity to elucidate the key high-order interactions (couplings) among features governing the target metrics ($E_a$ and $A$), as demonstrated in the section "**Important Interaction Search**" in the **Supplementary Information**. **These results collectively establish that the hierarchical inductive bias embedded in *GoodRegressor* constitutes a necessary structural principle for interpretable** **scientific AI in high-dimensional compositional spaces.** Furthermore, *GoodRegressor* exhibits excellent reproducibility in both regression performance and materials prediction through the deployment of stacking-ensembled models (see the sections "**Reproducibility and Promising Materials for Oxygen-ion Conductors"** and **"Materials Predictions"** in the **Supplementary Information**).

## Further Testbed Case Studies

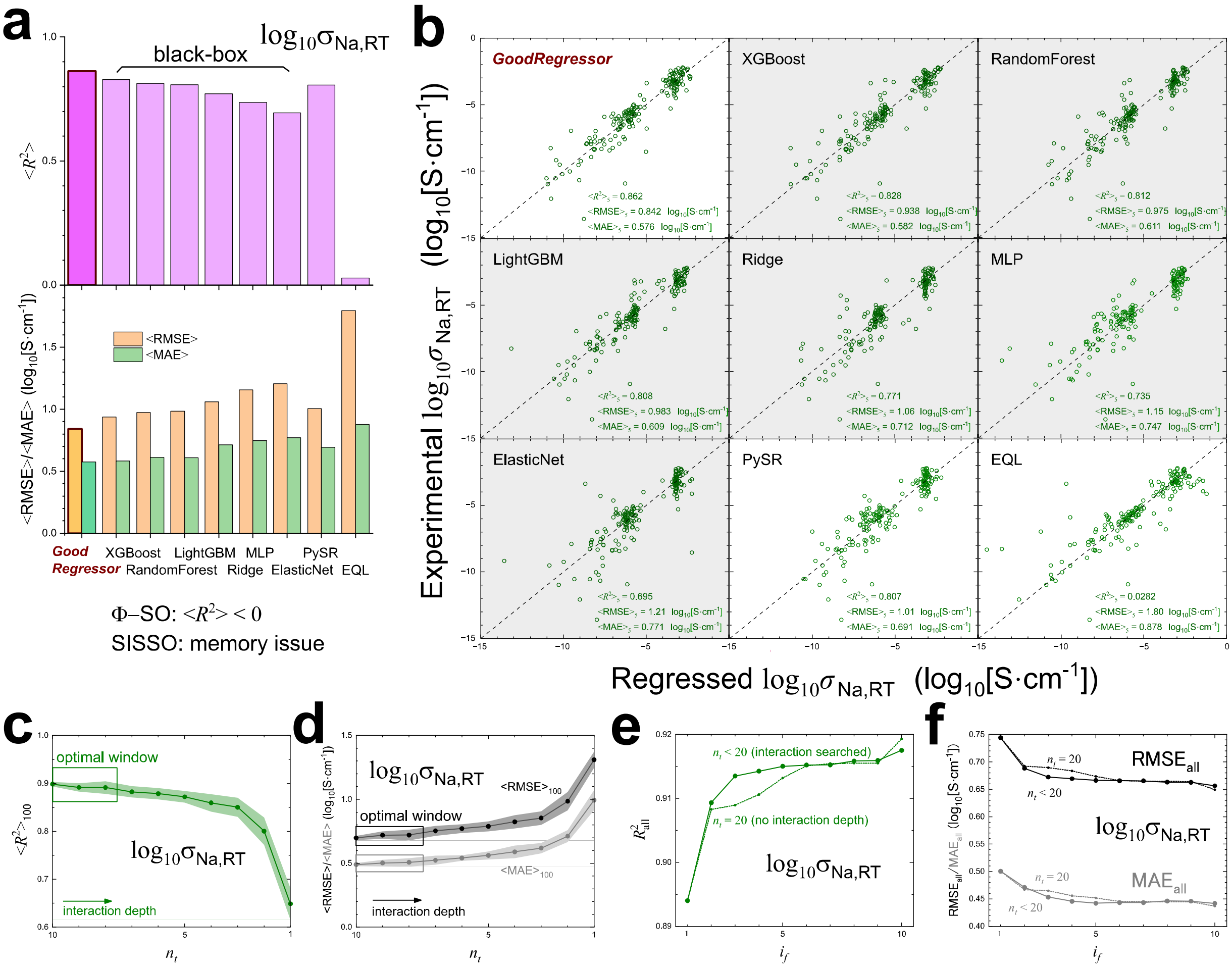

**Fig. 4** Benchmark comparison and hierarchical interaction analysis for room-temperature Na-ion conductivity $\sigma_{\mathrm{Na,RT}}$ of NASICONs (Na-ion super ionic conductors). (a) Comparison of predictive performance across *GoodRegressor* and representative machine learning models. (b) Experimental versus regressed parity plots for *GoodRegressor* and selected baselines. (c) Evolution of cross-validated $\langle R^2 \rangle_{100}$ with increasing hierarchical interaction depth, highlighting the optimal interaction window. (d) Corresponding evolution of $\langle \mathrm{RMSE} \rangle_{100}$ and $\langle \mathrm{MAE} \rangle_{100}$. (e) Saturation behavior of $R^2_{\mathrm{all}}$ for stacking-ensembled models $\overline{M_{f,\iota_f}}$ with increasing $i_f$, compared with the interaction-disabled baseline ($n_t = 20$, dashed line). (f) Corresponding evolution of error metrics

($\mathrm{RMSE_{all}}$ and $\mathrm{MAE_{all}}$) demonstrating convergence and robustness. The notations in panels (a, b) follow those used in **Fig. 2**, whereas those in panels (c-f) follow **Fig. 3**.

*GoodRegressor* can also be applied to other science problems. **Fig. 4a** presents the 5-fold regression benchmark results for the room-temperature Na-ion conductivity $\sigma_{\mathrm{Na,RT}}$ of NASICONs, using the manually curated dataset from Ref. 13. The dataset consists of $n_{\mathrm{data}} = 180$ and $n(\mathbf{X_1}) = 211$. $\mathbf{X_1}$ was constructed from the quantities $\chi - \chi_{\mathrm{O}}$, $M$, $Z$, $r_{\mathrm{VI}}$, $\rho$, $\rho_{\mathrm{mo}}$ , $\eta_f$, $B$, $G$, $\nu$, $\kappa$, $\alpha$, $\theta_D$, $n_c$, $r$, $\chi_{\mathrm{mol}}$, and $n_{\mathrm{Na}}$, accompanied by $\langle\cdots\rangle$, $\sigma(\cdots)$, and $r(\cdots)$, where $\chi_{\mathrm{mol}}$ and $n_{\mathrm{Na}}$ denote the molar magnetic susceptibility and Na-ion content, respectively; for the remaining descriptors, see **Supplementary Table 1**. Given $n_s = 109$ and $n_t = 10$ in addition to $n(\mathbf{X_1})$, the total number of possible regression models reaches $N_3^{\vee} = 5.99 \times 10^{124}$.

Despite the immense size of $N_3^{\vee}$, the *GoodRegressor* model clearly outperforms the other machine learning methods. Specifically, it achieved $\langle R^2\rangle_5 = 0.862$, $\langle\mathrm{RMSE}\rangle_5 = 0.842\ \log_{10}[\mathrm{S\cdot cm^{-1}}]$, and $\langle\mathrm{MAE}\rangle_5 = 0.576\ \log_{10}[\mathrm{K\cdot S\cdot cm^{-1}}]$ . The other models recorded lower performance ($\langle R^2\rangle_5 \leq 0.828$, $\langle\mathrm{RMSE}\rangle_5 \geq 0.938\ \log_{10}[\mathrm{S\cdot cm^{-1}}]$, and $\langle\mathrm{MAE}\rangle_5 \geq 0.582\ \log_{10}[\mathrm{S\cdot cm^{-1}}]$). The performance ranking among these was as follows: XGBoost, RandomForest, LightGBM, PySR, Ridge, MLP, ElasticNet, and EQL. The benchmark results for Φ-SO and SISSO are not presented for the same reasons outlined earlier. The parity plots in **Fig. 4b** further confirm that the *GoodRegressor* model provides excellent predictive accuracy as well.

Next, a new stacking-ensembled model was generated with an 8: 2 train-validation split of the full dataset, without reserving a separate test set, which achieved $R^2_{\mathrm{all}} = 0.918$, $\mathrm{RMSE_{all}} =$ $0.657 \ \log_{10}[\mathrm{S} \cdot \mathrm{cm}^{-1}]$, and $\mathrm{MAE}_{\mathrm{all}} = 0.442 \ \log_{10}[\mathrm{S} \cdot \mathrm{cm}^{-1}]$. **Figs. 4c and 4d** show the evolution of predictive performance for the NASICON dataset as hierarchical interaction depth increases. Unlike the oxygen-ion conductor case, the optimal performance is already achieved at $n_t \sim 20$, indicating that deeper interaction expansion provides only marginal improvement. The corresponding error metrics exhibit similarly weak dependence on interaction depth. This behavior suggests that the descriptor relationships governing $\sigma_{\mathrm{Na,RT}}$ in NASICONs are comparatively less hierarchically entangled. The stacking results in **Figs. 4e and 4f** further support this interpretation. Fixing $n_t = 20$ yields performance comparable to that obtained with interaction-enabled models, and the dashed baseline does not exhibit substantial degradation. This contrasts with the oxygen-ion conductor case, where disabling interaction construction resulted in clear performance loss. **Together, these findings indicate that while hierarchical interaction search can provide incremental gains, the NASICON problem is structurally more tractable and less dependent on deep compositional expansion. Consistently, heuristic symbolic methods such as PySR achieve performance comparable to *GoodRegressor* for this dataset, reflecting the relatively low degree of hierarchical descriptor entanglement.** Furthermore, a promising materials candidate is shown in the section **“Promising Material for NASICONs”** in the **Supplementary Information**.

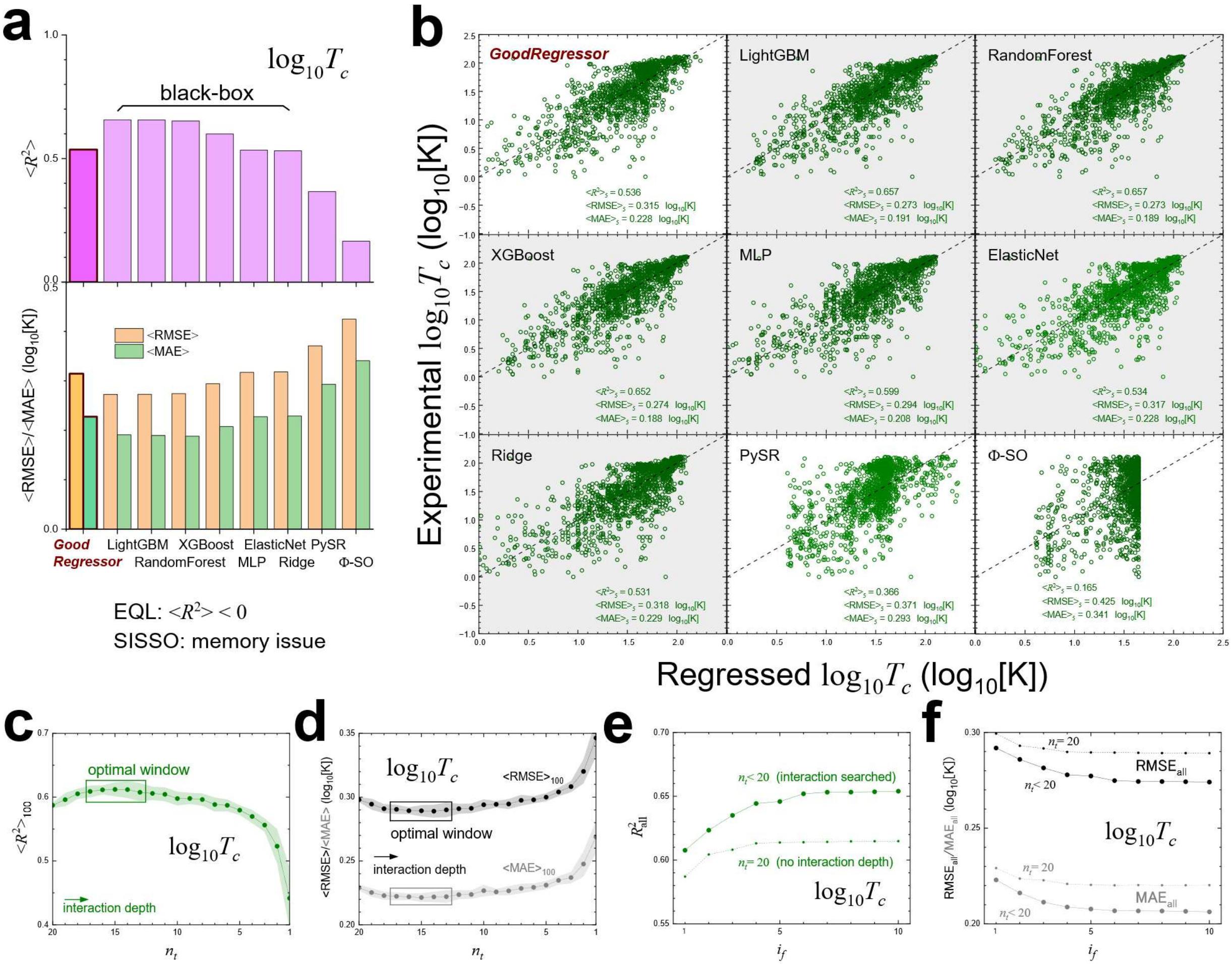

**Fig. 5** Benchmark and hierarchical interaction analysis for superconducting oxides. Panel structure and notations follow Fig. 8. In contrast to the NASICON case, a distinct optimal interaction window and substantial performance loss under interaction ablation are observed.

**Fig. 5a** presents the 5-fold regression benchmark results for the superconducting transition temperature $T_c$ of superconducting oxides, using the dataset from Ref. 14. Although the original dataset contained $n_{\text{data}} = 11{,}964$ entries, its size was reduced to $n_{\text{data}} = 1{,}358$ by removing compositions that were similar to higher-$T_c$ counterparts to reduce bias from correlated samples, improve model generalization with more informative dataset, and reduce training time. Specifically, if a composition $A_a B_b C_c D_d \mathrm{O}_o$ exhibited a lower $T_c$ than a composition $A_{a'} B_{b'} C_{c'} D_{d'} \mathrm{O}_{o'}$ with $|\mathcal{N}(a,b,c,d) - \mathcal{N}(a',b',c',d')| < 0.1$ where $\mathcal{N}$ denotes the normalized composition vector, then the lower-$T_c$ composition was discarded. This filtering procedure is implemented within the curator module. $\mathbf{X_1}$ was constructed from the quantities $\chi - \chi_{\mathrm{O}}$, $M$, $Z$, $r_{\mathrm{VI}}$, $\rho$, $\rho_{\mathrm{mol}}$, $\eta_f$, $B$, $G$, $\nu$, $\kappa$, $\alpha$, $\theta_D$, $n_c$, $r$, $\chi_{\mathrm{mol}}$, $n$, $l$, $U$, $t_{\mathrm{c}}$, and $\eta_{fueh}$, accompanied by $\langle \cdots \rangle$, $\sigma(\cdots)$, and $r(\cdots)$, where $U$, $t_{\mathrm{c}}$, and $\eta_{fueh}$ denote Hubbard parameters for transition-metal and rare-earth atoms,[15, 16] the superconducting transition temperatures for elemental metals, and the filling rate per unquenched electron or hole under the assumption of an octahedral crystal field (i.e., $\eta_{fueh} = \eta_f / N_{ueh}$, where $N_{ueh}$ is the number of unquenched electrons or holes, for example, $N_{ueh} = 1$ for $5d\ t_{2g}^5$), respectively; for the remaining descriptors, see **Supplementary Table 1**. Given $n_s = 109$ and $n_t = 20$ in addition to $n(\mathbf{X_1})$, the total number of possible regression models reaches $N_3^{\mathrm{V}} = 4.20 \times 10^{430}$.

Despite the immense size of $N_3^{\mathrm{V}}$, the *GoodRegressor* model's performance was comparable to state-of-the-art black box models and outperforms the white-box baselines. Specifically, it achieved $\langle R^2 \rangle_5 = 0.536$, $\langle \mathrm{RMSE} \rangle_5 = 0.315\ \log_{10}[\mathrm{K}]$, and $\langle \mathrm{MAE} \rangle_5 = 0.228\ \log_{10}[\mathrm{K}]$, despite **the absence of crystal-structure information.** The other models recorded $\langle R^2 \rangle_5 \leq 0.657$, $\langle \mathrm{RMSE} \rangle_5 \geq 0.273\ \log_{10}[\mathrm{S \cdot cm^{-1}}]$, and $\langle \mathrm{MAE} \rangle_5 \geq 0.188\ \log_{10}[\mathrm{S \cdot cm^{-1}}]$. The performance ranking among these was as follows: LightGBM, RandomForest, XGBoost, MLP, ElasticNet, Ridge, PySR, and Φ-SO. The benchmark results for EQL and SISSO are not presented for the same reasons outlined earlier. The parity plots in **Fig. 5b** further confirm that it provides considerable predictive accuracy as well.

Next, a new stacking-ensembled model was generated with an 8: 2 train-validation split of the full dataset, without reserving a separate test set, which achieved $R^2_{\text{all}} = 0.654$ , $\text{RMSE}_{\text{all}} = 0.274\ \log_{10}[\text{S} \cdot \text{cm}^{-1}]$ , and $\text{MAE}_{\text{all}} = 0.206\ \log_{10}[\text{S} \cdot \text{cm}^{-1}]$ . **Figs. 5c and 5d** display the evolution of predictive performance for the superconducting oxide dataset as hierarchical interaction depth increases. In contrast to the NASICON case, a pronounced optimal interaction window is clearly observed. The cross-validated $\langle R^2 \rangle_{100}$ increases substantially as interaction depth expands, reaching a maximum within a finite range ($17 \geq n_t \geq 13$) before declining at excessive compositional complexity, representing a broadened peak feature. The corresponding $\langle \text{RMSE} \rangle_{100}$ and $\langle \text{MAE} \rangle_{100}$ exhibit complementary minima within the same window. This behavior indicates strong hierarchical descriptor entanglement governing superconducting transition temperatures. The stacking evolution in **Figs. 5e and 5f** further reinforces this conclusion. As the number of stacked models $i_f$ increases, predictive performance improves and then saturates. Importantly, the dashed-line baseline corresponding to interaction-disabled models ($n_t = 20$) shows substantial performance degradation. Unlike the NASICON case, disabling hierarchical interaction construction here results in a marked reduction in $R^2_{\text{all}}$ and elevated error metrics. **These observations demonstrate that hierarchical interaction expansion is not merely beneficial but structurally essential for accurate modeling of superconducting systems.** Furthermore, a promising materials candidate is shown in the section **"Promising Material for Superconducting Oxides"** in the **Supplementary Information**.

Taken together, comparison across the three systems reveals distinct interaction-depth signatures. Oxygen-ion conductors exhibit a sharply defined optimal interaction window at $n_t < 20$ , indicating tightly tuned hierarchical entanglement in which predictive performance is highly sensitive to compositional depth. NASICON conductivity, by contrast, displays a broad and shallow depth dependence with an optimum near $n_t \approx 20$, suggesting comparatively weak descriptor coupling and limited necessity for deep interaction construction. Superconducting oxides also favor $n_t < 20$ but exhibit a broadened peak in the optimal window, implying strong yet partially redundant hierarchical entanglement.

**These patterns indicate that interaction-depth sensitivity encodes structural information about the underlying physical phenomenon.** Rather than merely distinguishing between "simple" and "complex" datasets, interaction-depth evolution reflects how compositional complexity is organized hierarchically. Three observable quantities characterize this organization: **the optimal depth location, the sharpness of the optimal window, and the magnitude of performance degradation when hierarchical interaction construction is disabled. Together, these quantities provide an empirical taxonomy of hierarchical complexity in scientific datasets.** Interaction depth therefore functions not only as a hyperparameter but as a structural probe. Systems with sharply tuned optima reflect tightly coupled descriptor entanglement, whereas broad or depth-insensitive regimes indicate weaker or partially redundant compositional structure. It is also noteworthy that *GoodRegressor* has successfully yielded a chemically sensible theoretical framework for metal hydrides.[17]

## DISCUSSION

This work advances hierarchical inductive bias with explicit depth control as a structural design principle for interpretable scientific AI. Predictive robustness emerges from disciplined compositional expansion governed by controlled interaction depth. The results demonstrate that predictive performance follows a system-dependent interaction-depth evolution, exhibiting characteristic optima that balance structural entanglement against over-complexity. **Unlike conventional hyperparameters that primarily regulate model capacity, interaction depth in *GoodRegressor* governs the structural grammar by which scientific hypotheses are constructed.**

Across oxygen-ion conductors, NASICONs, and superconducting oxides, interaction-depth sensitivity reveals distinct structural signatures. These signatures are reproducible under ensemble averaging, addressing concerns regarding combinatorial explosion and structural instability in symbolic regression. Hierarchical depth control therefore serves both as a regularization mechanism and as a structural probe of descriptor organization. **Although demonstrated on materials datasets, the principle is domain-agnostic: any scientific system characterized by hierarchical descriptor entanglement is expected to exhibit an interaction-depth signature. From a computational science perspective, interaction depth therefore provides a general diagnostic axis for probing hierarchical complexity in such systems, including domains such as chemistry, biology, and complex physical systems where compositional interactions govern emergent behavior.**

**Importantly, the findings suggest that scientific datasets differ not merely in predictive difficulty but in how compositional complexity is organized.** The location of the optimal depth,

the sharpness of its window, and the magnitude of degradation when interactions are disabled together define an empirical taxonomy of hierarchical complexity in high-dimensional compositional spaces. This perspective reframes symbolic regression from a purely algorithmic exercise into a question of inductive-bias design. When physical targets arise from nonlinear and multi-level descriptor entanglement, structured hierarchical construction is not optional but necessary. Hierarchical depth-controlled symbolic exploration provides a principled pathway toward interpretable AI in deeply entangled scientific systems. **By unifying interpretability, scalability, reproducibility, and structural diagnostics, this framework of *GoodRegressor* suggests a path toward scientific AI systems that do not merely approximate data but reveal how hierarchical complexity governs physical phenomena.**

## METHODS

### Symbolic Regression Algorithm

#### a. Run through (the jungle)

As illustrated in **the section "Symbolic Regression Algorithm: a Structure Loosely Reminiscent of Terence McKenna's *Timewave Zero*"** in the **Supplementary Information**, given the number of elements, that is, $n(\mathbf{X_1})$, for a specified number $N_t$ of active variables, the number of all possible combinations are given as $\binom{n(\mathbf{X_1})}{n_t}$ (see **Supplementary Fig. 10a**). In this study, $n(\mathbf{X_1}) = 358$ and $n_t = 20$ were taken, yielding $\binom{n(\mathbf{X_1})}{n_t} \cong 2.86 \times 10^{32}$. However, because the search space can be astronomically large, it is divided lexicographically and distributed across $n_c$ CPU cores (herein, $n_c \geq 2048$). Each core samples the ordered model space at a fixed interval $n_j$, named "jumping-jack-flash" interval (e.g., evaluating the 1st, $(1 + n_j)$-th, $(1 + 2n_j)$ models, etc.), where $n_j$ is tuned to satisfy a predefined computational time limit for this step, "run through (the jungle)" (herein, $1000$ sec). This can be achieved by directly identifying the ∃-th combination in the lexicographic order, rather than iteratively updating combinations up to ∃ (as in, for example, std::next_permutation in C++). The detailed implementation is provided in the section **"Search Code in the Lexicographic Order"** in the **Supplementary Information**.

Within each core, the model with the highest coefficient $R^2_{\mathrm{vald}}$ of determination of the validation dataset that satisfies the "full Fisher" condition (or "full frequentist" condition) $F_f$ (strict $p$-value constrains), namely, $p(F) < 0.05$ in the $F$-test and $p(t) < 0.05$ in the $t$-tests for all coefficients and the intercept is retained as the provisional best model: $M_{i_c}$ for the $i_c$-th core.

It is noted that the maximum possible model index, corresponding to the upper limit of the model search line number, that is, $\binom{n(\mathbf{X_1})}{n_t}$ is $10^{4932}$ (or $10^{308}$ under MSVC on Windows), which equals the maximum representable value of a long double variable in the C++ compiler used. This defines the theoretical upper bound of the regression model space that the algorithm can reference. This core component was developed on the basis of the *EwaldSolidSolution* code, originally implemented to rapidly determine the global site configurations of ionic solid solutions.[18]

**b. Swap**

Starting from each core's best model $M_{i_c}$, a local refinement step ("swap") is executed (see **Supplementary Fig. 10b**). The variable with the largest $p$-value (least statistically significant) is temporarily removed and replaced, one by one, with currently inactive variables. Each swapped model is evaluated under $F_f$, and the configuration yielding the highest $R^2_{\text{vald}}$ is retained. This procedure is repeated from the least (largest $p$) to the most significant variable (smallest $p$), allowing the algorithm to refine models by exploring regions of the model space not examined during the initial sampling.

**c. Transit**

To capture nonlinear effects, the algorithm next applies scalar transformations to the active variables (see **Supplementary Fig. 10c**). Beginning with the most significant variable (smallest $p$-value), various transformations (e.g., $\text{erf}(x_{i_i})$, $\log(x_{i_i})$, $\exp(x_{i_i})$, $\sin(x_{i_i})$, $\sqrt{x_{i_i}}$, $x_{i_i}^2$, $\cdots$) are tested (herein, 109 transformations; provided in the section "**Scalar Transform List**" in the **Supplementary Information**). Each transformed model is evaluated under $F_f$, and the form giving the highest $R^2_{\text{vald}}$ is selected. This process is repeated sequentially for the remaining variables in order of increasing $p$-value. The swap and transit steps are alternated until $R^2_{\text{vald}}$ converges, ensuring a statistically and numerically optimized local solution.

**d. Pick**

After completing the **a–c** sequence on all cores, the model with the highest $R^2_{\text{vald}}$ across cores is selected as the current global optimum (see **Supplementary Fig. 10d**). The algorithm then constructs a new model with $(n_t \rightarrow) n_t - 1$ active variables, but from an expanded candidate pool that is larger than the original feature space. This candidate pool set $\mathbf{X}_i$ $(i = 2, 3, \cdots)$ is given as the union of the original descriptor set $\mathbf{X_1} = \mathbf{X}_1^{\circ} \cup \mathbf{X}_1^{m} \cup \mathbf{X}_1^{d}$, the scalar-transformed variable set given by the $i$-th global optimum $\mathbf{X}_i^{\circ}$ (e.g., $\mathrm{erf}(x_{i_i})$, $\log(x_{i_i})$, $\exp(x_{i_i})$, $\sin(x_{i_i})$, $\sqrt{x_{i_i}}$, $x_{i_i}^2$, $\cdots$), its multiplication interaction set $\mathbf{X}_i^{m}$ (e.g., $\mathrm{erf}(x_{i_i}) \exp\left(x_{i_j}\right)$, $\cdots$), and its division interaction set (e.g., $\mathrm{erf}(x_{i_i}) / \exp\left(x_{i_j}\right)$, $\cdots$). Thus, even though the number of active variables is reduced by one (i.e., $n_t - 1$), the search space itself becomes richer and more expressive, incorporating nonlinear and cross-transformed combinations. From this expanded pool, the run through, swap, and transit cycles are repeated under $F_f$ until $\langle R^2 \rangle_{100}$ no longer improves. The final expression obtained from this procedure is denoted $M_f$, representing a statistically validated, parsimonious symbolic model. It is noteworthy that a simple regression model can be obtained by performing only a single run (i.e., without applying the "pick" procedure), which may be useful when such a straightforward model sufficiently meets the research objectives.

**e. Bagging**

To enhance model robustness and mitigate overfitting, the entire pipeline (**a–d**) is repeated $n_f$ times (typically $n_f = 10$ iterations) with different train-validation splits (see **Supplementary Fig.**

**10e**). The $i_f$-th iteration yields an independent final model $M_{f,i_f}$, and ensemble averaging up to $M_{f,i_f}$ produces a consensus model $\overline{M_{f,i_f}}$ with converged overall $R^2_{\mathrm{all}}$ (applied to all the data points): finally, $\overline{M_{f,n_f}}$ . As the number of ensemble members increases, both the mean $R^2_{\mathrm{all}}$ and the partial dependence plots stabilize, indicating improved statistical robustness and reduced risk of overfitting or noise sensitivity. The $R^2_{\mathrm{all}}$ values converged within ten iterations, indicating that further repetitions did not significantly improve model performance. The final ensemble-averaged symbolic model thus provides a statistically rigorous and physically interpretable representation of the relationship between activation energy and its underlying descriptors.

**Associated content**

Supplementary Information: work flow, scalar transform list, details of conventional machine learning approaches, models $M_{f,1}$ for $E_a$ and $A$, full descriptions of $M_{f,1}$ for $E_a$ and $A$, important interaction search, reproducibility and promising materials for oxygen-ion conductors, materials predictions, promising material for NASICONs, promising materials for superconducting oxides, symbolic regression algorithm: a structure loosely reminiscent of Terence McKenna's *Timewave Zero*, and search code in the lexicographic order

**Author information**

**Code Availability**

The source codes of *GoodRegressor* and benchmark tests are available at https://github.com/JerryGarcia1995/OxygenIonConductor.

**Acknowledgments**

The authors thank Prof. Yu Kumagai and Dr. Soungmin Bae in Tohoku University and Mr. Seongcheol Jeong in LG Innotek for informative discussions. Parts of the numerical calculations have been done using the facilities of the Supercomputer Center, the Institute for Solid State Physics, University of Tokyo.

Supplementary Information

# *GoodRegressor*: A Hierarchical Inductive Bias for Navigating High-Dimensional Compositional Space

*Seong-Hoon Jang*[*1]

[1] Institute for Materials Research (IMR), Tohoku University, Sendai, 980-8577

*Corresponding author: jang.seonghoon.b4@tohoku.ac.jp (S.-H. Jang)

This file contains:

## Workflow

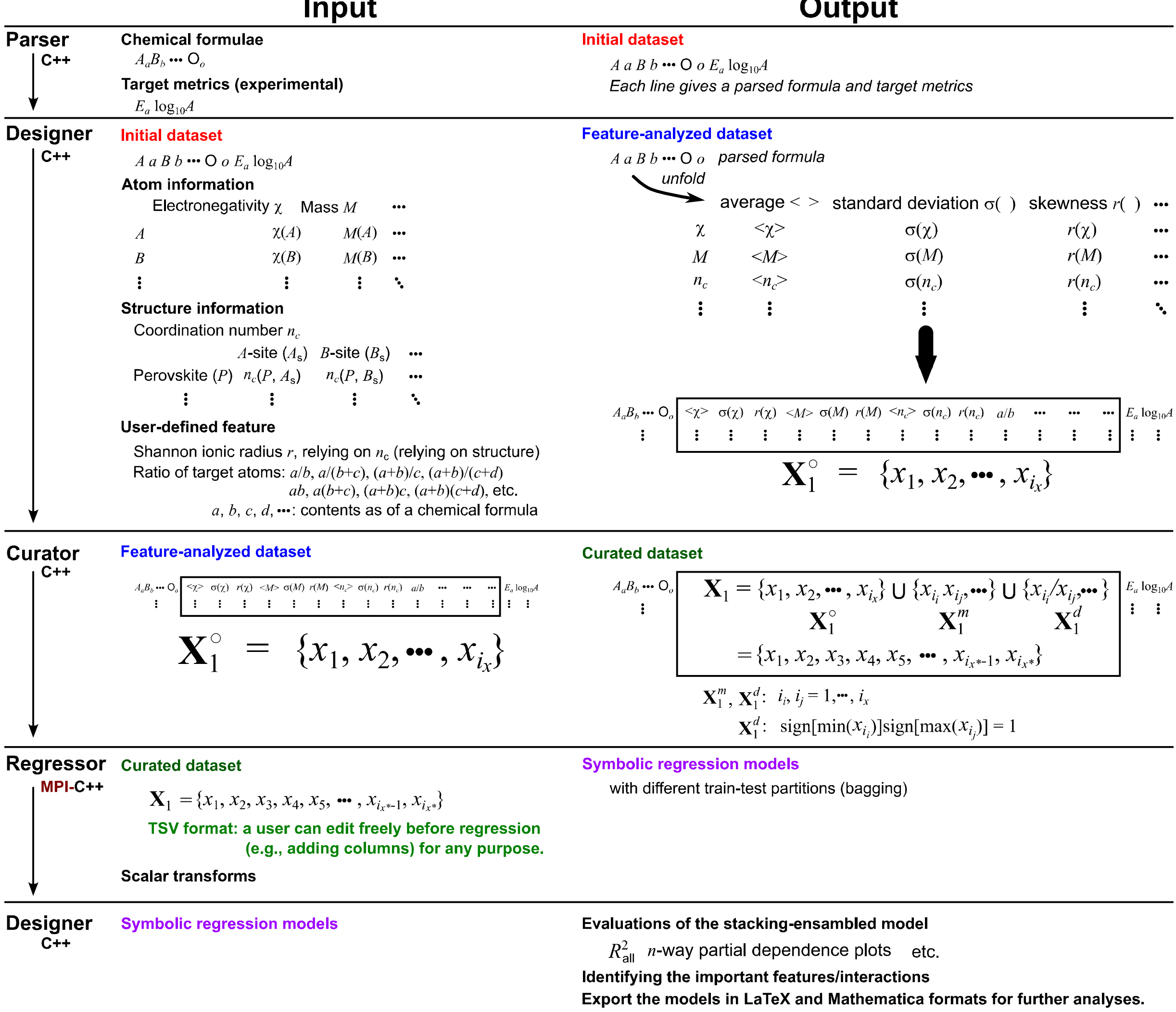

**Supplementary Fig. 1** Workflow of *GoodRegressor*. The framework comprises multiple steps, parser, designer, curator, regressor, and designer (as a post-processing step) modules, with their corresponding inputs and outputs explicitly indicated. The parser module produces parsed chemical formulae paired with target metrics. The designer module generates a feature-analyzed dataset by calculating statistical descriptors such as the average, standard deviation, and skewness of the selected features (see **Supplementary Table 1**). The curator module constructs interaction

features by computing the products and ratios of feature values, which are combined with the basic feature set obtained from the designer. Using the resulting curated dataset, the regressor module performs symbolic regression to build predictive models. This enables ensemble model evaluation (via stacking ensemble), identification of key features and feature interactions associated with the target metrics, and further materials predictions with the designer module (as a post-process).

**Supplementary Table 1** Structural, chemical, and physical properties of constituent elements of oxides for regression models, given as "features" for symbolic regression modeling.

| Properties | Description | Unit |
|---|---|---|
| $\bar{O}$ | Molar ratio of oxygen ions to metal ions | - |
| $M$ | Atomic mass | $\mathrm{g \cdot mol^{-1}}$ |
| $Z$ | Valence | - |
| $r_{\mathrm{VI}}$ | Shannon ionic radius with sixfold coordination to oxygen[13, 14] | Å |
| $r$ | Shannon ionic radius depending on $n_c$ (see below)[S2, S3] | Å |
| $B$ | Bulk modulus | GPa |
| $G$ | Shear modulus | GPa |
| $\rho$ | Density | $\mathrm{g \cdot cm^{-3}}$ |
| $\rho_{\mathrm{mol}}$ | Molar density, defined as $\rho/M$ | $\mathrm{mol \cdot cm^{-3}}$ |
| $\eta_f$ | Ionic filling rate (per unit volume), defined as $\eta_f = N_{\mathrm{avo}}\rho_{\mathrm{mol}}\left(\frac{4}{3}\pi r_{\mathrm{VI}}^3\right)$ where $N_{\mathrm{avo}}$ is the Avogadro's constant | - |
| $n$ | Principal quantum number of valence electrons | - |
| $l$ | Azimuthal quantum number of valence electrons | - |
| $\alpha$ | Thermal expansion coefficient (linear not volumetric) | $\mathrm{K^{-1}}$ |
| $\kappa$ | Thermal conductivity | $\mathrm{W \cdot m^{-1} \cdot K^{-1}}$ |
| $\chi - \chi_\mathrm{O}$ | Difference in electronegativity between metal ions and oxygen | - |
| $\nu$ | Poisson's ratio | - |
| $\theta_D$ | Debye temperature | K |
| $n_c$ | Coordination number to oxygen ions, which depends on the occupied sites ($A_s$, $B_s$, or $C_s$) in different crystal classes. For example, in perovskite oxides, $n_c$ for $A_s$ and $B_s$ are given as 12 and 6, respectively. | - |
| $\langle \cdots \rangle$ | Average over the constituent metal ions | Common to $\cdots$ |
| $\sigma(\cdots)$ | Standard deviation over the constituent metal ions | Common to $\cdots$ |
| $r(\cdots)$ | Skewness over the constituent metal ions | - |

**Supplementary Fig. 1** illustrates the workflow of *GoodRegressor*, showing the inputs and outputs for each component. The overall workflow comprises five main steps: parser, designer, curator, regressor, and designer (used again as a post-processing step) modules. All components are implemented in C++, while the regressor module is computationally optimized through the implementation of the Message Passing Interface (MPI). To demonstrate the workflow in a concrete setting, I begin with the first case study: an oxygen-ion conductor database comprising $n_{\mathrm{data}} = 483$ samples,[S1] characterized by activation energies ($E_a$) and Arrhenius prefactors ($A$).

**a. Parser**

First, the parser takes as input the chemical formulae of a database along with target metrics serving as dependent variables ($E_a$ and $A$).[S1] The parser outputs an initial dataset consisting of parsed chemical formulae paired with the corresponding target metric values for each entry.

**b. Designer**

Second, this initial dataset is passed to the designer module. In this stage, additional input files (such as atomic information, structural information, and user-defined features) are also utilized. The atomic information file may include fundamental atomic properties such as electronegativity ($\chi$) and atomic mass ($M$). The list of adopted features is presented in **Supplementary Table 1** (hereafter, feature symbols are used without further denotation). The structural information file contains features dependent on structural characteristics rather than atomic ones, such as $n_c$. Furthermore, user-defined features can be freely incorporated, which depends on structural parameters and atomic characteristics both (e.g., $r$ depending on $n_c$ and atom),[S2, S3] or on ratios/products involving specific atoms (e.g., $\bar{O}$). Based on these inputs, the designer computes statistical descriptors for each parsed chemical formula, namely, the average, standard deviation, and skewness of each feature (as well as minimum, maximum, and kurtosis values, though they are generally not recommended for subsequent modeling). These computed descriptors, together with user-defined feature values, form the feature-analyzed dataset, where each line contains the derived feature values alongside the target metrics. This basic feature set is denoted as $\mathbf{X}_\mathbf{1}^\circ = \{x_1, x_2, \cdots, x_{i_x}\}$.

**c. Curator**

Third, the curator module processes the feature-analyzed dataset to identify feature interactions, thereby producing the curated dataset. Specifically, it constructs the union $\mathbf{X_1}$ of $\mathring{\mathbf{X}}_\mathbf{1}$, its multiplication interaction set $\mathbf{X_1^{\boldsymbol{m}}} = \left\{x_{i_i}x_{i_j}, \cdots\right\}$, and the division interaction set $\mathbf{X_1^{\boldsymbol{d}}} = \left\{x_{i_i}/x_{i_j}, \cdots\right\}$. For the construction of $\boldsymbol{X_1^d}$, to ensure numerical stability in constructing $\boldsymbol{X_1^d}$, each variable $x_{i_i}$ and $x_{i_j}$ must satisfy $\mathrm{sgn}\left[\min\left(x_{i_i}\right)\right]\mathrm{sgn}\left[\max\left(x_{i_i}\right)\right] = 1 \wedge \mathrm{sgn}\left[\min\left(x_{i_j}\right)\right]\mathrm{sgn}\left[\max\left(x_{i_j}\right)\right] = 1$, respectively. Here, it is given that $n(\mathbf{X_1}) = 358$ for the oxygen-ion conductor tasks.

**d. Regressor**

Fourth, the curated dataset, output in TSV format, which can be freely edited, is fed into the regressor module. As described in the subsection **"Regressor Module: Symbolic Regression Algorithm"** in the **Methods** section, this module independently generates $N_f$ symbolic models $M_{f,i_f}$ by varying the **random** train-validation splits with the ratio of $8:2$ **(bagging)**, where $i_f$ denotes the iteration number ($i_f = 1, \ldots, N_f$). Given $n(\mathbf{X_1}) = 358$, the possible simple linear combinations obtained by selecting $n_t$ features amount to $N_1^{\vee} = \binom{n(\mathbf{X_1})}{n_t}$; for $n_t = 20$ as taken in this study, it is given that $N_1^{\vee} = 2.86 \times 10^{32}$. It is noteworthy that the upper limit of the model search number in simple linear combinations is the long double limit $10^{4932}$ (or $10^{308}$ under Microsoft Visual C++ or MSVC on Windows), which remains considerably higher than $N_1^{\vee}$. Meanwhile, incorporating $n_s$ scalar transformations and interaction terms allows the model search to extend over a much larger number of possible combinations. In this study, $n_s = 109$ is provided (see the section **"Scalar Transform List"** in the **Supplementary Information**), which yields the combination number of $N_2^{\vee} = N_1^{\vee} {n_s}^{n_t} = 1.60 \times 10^{73}$. The interaction terms, according to the algorithm, allows the model optimization among the combination number of $N_3^{\vee} = N_2^{\vee} +$ $\sum_{i=1}^{n_t-1} N_{3,i}^{\vee}$ with $N_{3,1}^{\vee} = N_2^{\vee}\binom{n(\mathbf{X_1}) + n_t + 2\binom{n_t}{2}}{n_t - 1}$ and $N_{3,i}^{\vee} = N_{3,i-1}^{\vee}\binom{n(\mathbf{X_1}) + n_t - i + 1 + 2\binom{n_t - i + 1}{2}}{n_t - i}$ for $2 \leq i \leq n_t - 1$: $N_3^{\vee} = 1.44 \times 10^{457}$. In this procedure, as the number of active variables is reduced stepwise (i.e., $n_t \rightarrow n_t - 1$), the hierarchical interaction depth correspondingly increases. This occurs because limiting additive components forces the model to rely on progressively higher-order composite interaction terms, thereby deepening the symbolic hierarchy (for technical details, see the subsection **"Regressor Module: Symbolic Regression Algorithm"** in the **Methods** section). During this evolution, each symbolic regression model is evaluated using $100$ independent random re-samplings, each drawing 20 % out of the whole dataset including the training and validation ones. The averaged performance metrics, $\langle R^2 \rangle_{100}$, $\langle \mathrm{RMSE} \rangle_{100}$, and $\langle \mathrm{MAE} \rangle_{100}$, are computed across these re-samplings. At a specific value of $n_t$, these metrics reach their optimal values. For each interaction-depth evolution, the model achieving the highest $\langle R^2 \rangle_{100}$ is selected.

**e. Designer (as a post-process)**

Finally, the designer is employed again as a post-processing tool for the symbolic regression results. Staking the ensemble of models up to iteration $i_f$ yields a consensus model $\overline{M_{f,i_f}} = m_0 + \sum_{k_f=1}^{i_f} m_{k_f} M_{f,k_f}$, where the constant term $m_0$ and the coefficients $m_{k_f}$ are determined by the least-squares method with respect to the target metrics ($E_a$ and $A$). The final consensus model is thus denoted as $\overline{M_{f,n_f}}$, which allows for comprehensive evaluation of performance metrics such as overall coefficient of determination ($R^2_{\mathrm{all}}$), the root mean square errors ($\mathrm{RMSE}_{\mathrm{all}}$), and the mean absolute errors ($\mathrm{MAE_{all}}$) for the “entire” dataset including train and validation data points, all of which converge as $i_f$ increases. Here, it is given that $n_f = 10$.

One of the advantages of employing symbolic regression modeling, a white-box approach, is its ability to reveal not only the important individual **features** but also the significant **interactions** among them. These interactions manifest as the co-occurrence of multiple features within a single term, and their number is referred to as the interaction level ($l_{\exists M}$). As described in the subsection ***“Designer Module as a Post-process: Identification of Important Interaction Chains”*** in the **Methods** section, the designer module (applied as a post-processing step) quantifies two key aspects: (i) the frequency of appearance ($n_{\exists M}$), representing how many models contain term(s) in which the target features coexist and (ii) the weighted-average coefficient magnitude ($z_{\exists M}$), defined as the mean of the absolute values of the $z$-scored coefficients for such terms across all generated models, weighted by $w_{i_f} = m_{i_f} / \sum_{k_f=1}^{N_f} \left| m_{k_f} \right|$. Given $\overline{M_{f,n_f}}$, it is also possible to do further materials predictions as described in the subsection **“Designer Module as a Post-process: Materials Predictions”** in the **Methods** section. In addition, this module exports the model equations in LaTeX and Mathematica formats to facilitate users’ further independent analyses.

**Scalar Transform List**

Polynomial series: $x$, $x^{-3}$, $x^{-2}$, $x^{-1}$, $x^{-1/2}$, $x^{-1/3}$, $x^{1/3}$, $x^{1/2}$, $x^2$, $x^3$

Logarithm series: $\log_{10}x$, $[\log_{10}x]^{-3}$, $[\log_{10}x]^{-2}$, $[\log_{10}x]^{-1}$, $[\log_{10}x]^2$, $[\log_{10}x]^3$

Power series: $10^x$, $10^{-3x}$, $10^{-2x}$, $10^{-x}$, $10^{2x}$, $10^{3x}$

Exponential series: $e^x$, $e^{-3x}$, $e^{-2x}$, $e^{-x}$, $e^{2x}$, $e^{3x}$

Error-function series: $\mathrm{erf}(x)$, $\mathrm{erf}(x/1000)$, $\mathrm{erf}(x/100)$, $\mathrm{erf}(x/10)$, $\mathrm{erf}(10x)$, $\mathrm{erf}(100x)$, $\mathrm{erf}(1000x)$, $[\mathrm{erf}(x)]^2$, $[\mathrm{erf}(x/1000)]^2$, $[\mathrm{erf}(x/100)]^2$, $[\mathrm{erf}(x/10)]^2$, $[\mathrm{erf}(10x)]^2$, $[\mathrm{erf}(100x)]^2$, $[\mathrm{erf}(1000x)]^2$, $\mathrm{erf}(x-1/20)$, $\mathrm{erf}(x-1/10)$, $\mathrm{erf}(x-1/2)$, $\mathrm{erf}(x-7/8)$, $\mathrm{erf}(x-1)$, $\mathrm{erf}(x-5)$, $\mathrm{erf}(x-10)$, $\mathrm{erf}(x-50)$, $\mathrm{erf}(x-100)$, $\mathrm{erf}(x-1000)$, $\mathrm{erf}(x-10000)$, $[\mathrm{erf}(x-1/20)]^2$, $[\mathrm{erf}(x-1/10)]^2$, $[\mathrm{erf}(x-1/2)]^2$, $[\mathrm{erf}(x-7/8)]^2$, $[\mathrm{erf}(x-1)]^2$, $[\mathrm{erf}(x-5)]^2$, $[\mathrm{erf}(x-10)]^2$, $[\mathrm{erf}(x-50)]^2$, $[\mathrm{erf}(x-100)]^2$, $[\mathrm{erf}(x-1000)]^2$, $[\mathrm{erf}(x-10000)]^2$, $[\mathrm{erf}(x-7/8)]^3$, $[\mathrm{erf}(x-7/8)]^4$

Sine series: $\sin(x)$, $\sin(\pi x/2)$, $\sin(\pi x)$, $\sin(x/1000)$, $\sin(\pi x/2000)$, $\sin(\pi x/1000)$, $\sin(x/100)$, $\sin(\pi x/200)$, $\sin(\pi x/100)$, $\sin(x/10)$, $\sin(\pi x/20)$, $\sin(\pi x/10)$, $\sin(10x)$, $\sin(5\pi x)$, $\sin(10\pi x)$, $\sin(100x)$, $\sin(50\pi x)$, $\sin(100\pi x)$, $\sin(1000x)$, $\sin(500\pi x)$, $\sin(1000\pi x)$

Cosine series: $\cos(x)$, $\cos(\pi x/2)$, $\cos(\pi x)$, $\cos(x/1000)$, $\cos(\pi x/2000)$, $\cos(\pi x/1000)$, $\cos(x/100)$, $\cos(\pi x/200)$, $\cos(\pi x/100)$, $\cos(x/10)$, $\cos(\pi x/20)$, $\cos(\pi x/10)$, $\cos(10x)$, $\cos(5\pi x)$, $\cos(10\pi x)$, $\cos(100x)$, $\cos(50\pi x)$, $\cos(100\pi x)$, $\cos(1000x)$, $\cos(500\pi x)$, $\cos(1000\pi x)$

**Details of Conventional Machine Learning Approaches**

To evaluate the predictive performance of various regression algorithms on the dataset, a standardized benchmarking pipeline was implemented in Python. This framework provides a uniform and unbiased comparison between conventional machine-learning and symbolic regression approaches. The evaluation follows a nested cross-validation design with systematic hyperparameter optimization, ensuring fair and reproducible comparison across models. Each model underwent 5-fold nested cross-validation. The details of parameters are given below.

**a. Ridge**

Core library: scikit-learn

Search parameters: $\alpha \in [10^{-4}, 10^{3}]$

Iterations: 80

**b. ElasticNet**

Core library: scikit-learn

Search parameters: $\alpha \in [10^{-4}, 10^{1}]$, $l_1$ ratio $\in [0, 1]$

Iterations: 80

**c. MLP**

Core library: scikit-learn

Search parameters: Hidden layers: (256, 128), (256, 128, 64), (512, 256, 128); activation: ReLU or tanh; $\alpha \in [10^{-6}, 10^{-2}]$; learning rate $\in [3 \times 10^{-4}, 3 \times 10^{-2}]$; max_iter = 4000

Iterations: 80

**d. RandomForest**

Core library: scikit-learn

Search parameters: n_estimators ∈ [800, 2000], max_depth ∈ [6, 28], min_samples_split ∈ [2, 12], min_samples_leaf ∈ [1, 6], max_features ∈ [0.3, 0.7]

Iterations: 100

**e. XGBoost**

Core library: xgboost

Search parameters: n_estimators ∈ [1200, 3000], learning rate ∈ [0.01, 0.2], max_depth ∈ [3, 12], subsample ∈ [0.6, 1.0], reg_lambda ∈ [1, 80], gamma $\in [10^{-9}, 10^{-1}]$

Iteration: 140

**f. LightGBM**

Core library: lightgbm

Search parameters: n_estimators ∈ [1500, 4000], learning rate ∈ [0.01, 0.2], num_leaves ∈ [31, 255], min_child_samples ∈ [5, 120], feature/bagging fraction ∈ [0.6, 1.0], $\lambda_1$, $\lambda_2 \in [10^{-3},10]$

Iteration: 140

**g. EQL**

Epochs: 600

Learning rate: $1 \times 10^{-3}$

$L_1$ penalty: $1 \times 10^{-3}$ on output weights and projection layers

Activation functions: {sin, cos, exp, log, erf, square, cube, linear, multiplicative interactions}

Term constraint: maximum 20 active symbolic terms via adaptive top-K masking

**h. PySR**

Iterations: 220

Populations: 12

Maximum Expression Size: 40

Operators: ninary: {+, −, ×, ÷, *pow*}; unary: {exp, log, sin, cos, tan, sqrt, abs}

Loss Function: $L_2$ distance

Model Selection: best expression by validation loss

**i. Φ-SO**

Each symbolic search ran for 60 epochs, operating on symbolic operators {mul, add, sub, div, $n^2$, sqrt, neg, exp, log, sin, cos}.

## Models $M_{f,1}$ for $E_a$ and $A$

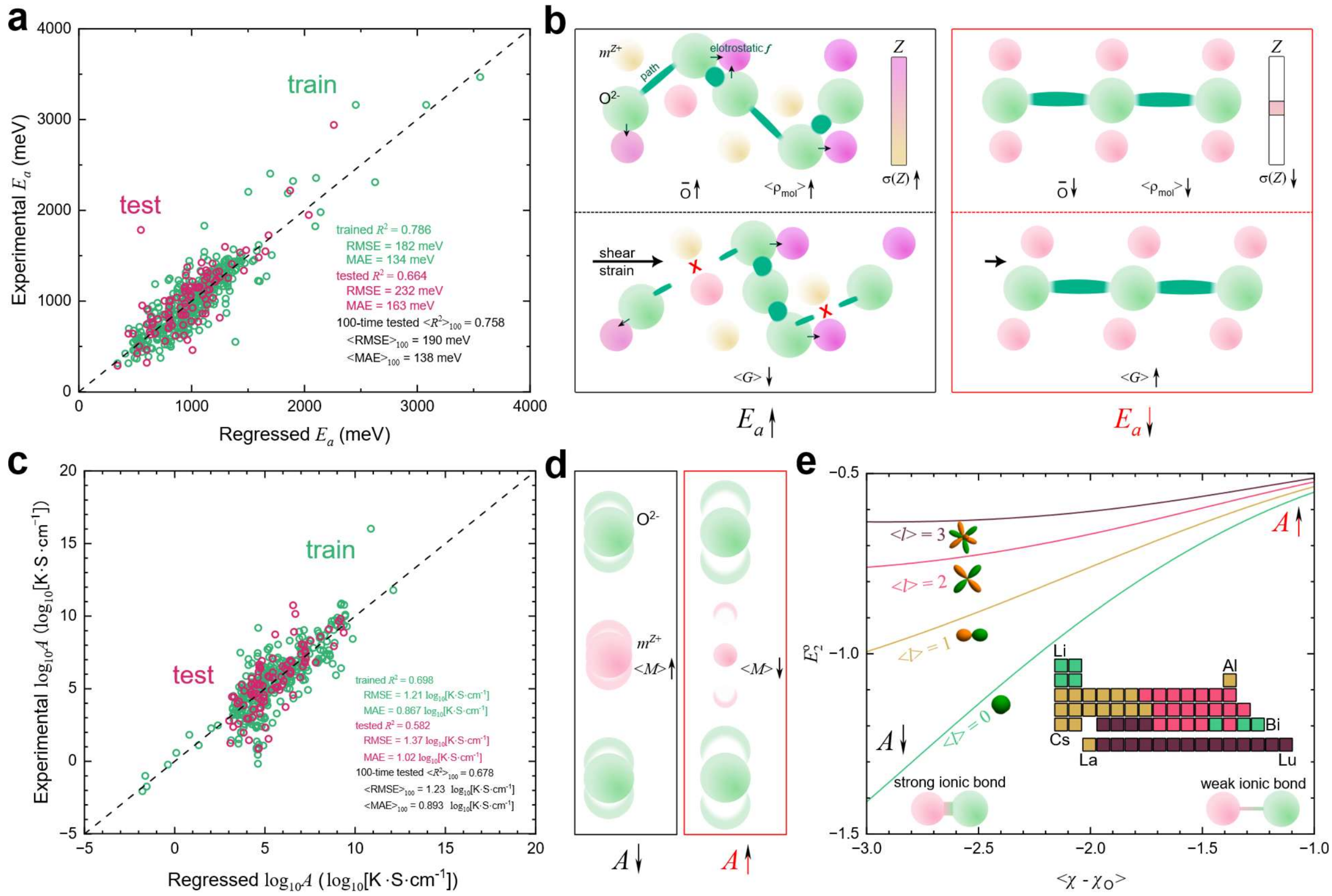

**Supplementary Fig. 2** Symbolic regression models $M_{f,1}$ and major descriptors for oxygen ion conductivity. (a) Parity plot for the activation energy $E_a$ model. (b) Representative structural and physical descriptors contributing to $E_a$, highlighting the effects of low charge disorder [$\sigma(Z)$], low oxygen rate ($\bar{O}$), loose packing density (small $\langle \rho_{\mathrm{mol}} \rangle$), and rigid shear modulus (high $\langle G \rangle$) on lowering $E_a$. (c) Parity plot for the $\log_{10} A$ model. (d) Low average atomic mass as a key descriptor for enhancing $A$. (e) Other key descriptors for $A$ including low electronegativity differences (small size of $\langle \chi - \chi_{\mathrm{O}} \rangle$ or high $\langle \chi \rangle$) and high orbital anisotropy ($\langle l \rangle$), which collectively enhance $A$ through the term $E_2^{\circ}$. For the illustrative plot of $E_2^{\circ}$, the parameters $\langle r_{\mathrm{VI}} \rangle \langle \rho_{\mathrm{mol}} \rangle = 0.07$, $B_2 = 0.7$,

and $\sigma(G) = 0$ were adopted for simplicity. The inset illustrates the value of $l$ assigned to each metal ion in the periodic table, with colors corresponding to the line colors in the main plot.

**Each iteration model $M_{f,i_f}$ conveys its own narrative of the physics embedded in the target variable.** Yet, given the intricate interrelations within $E_a$ and $\log_{10} A$, interpretation through a single $M_{f,i_f}$ risks oversimplifying the underlying complexity. Echoing Hans-Georg Gadamer's concept of the *fusion of horizons*, each model engages in a hermeneutical conversation with the others, together enriching the collective understanding of the physical system. As an example, the physical hermeneutics of the first-achieved models $M_{f,1}$ for $E_a$ and $\log_{10} A$ will be explained below.

The prediction capability of an "individual" model $M_{f,1}$ for $E_a$ is shown in **Supplementary Fig. 22a**. The model satisfied the $F$-test with $p$-value less than $10^{-2}$ , and the coefficient ($c_i$) of each descriptor ($x_i$) and the intercept term ($c_0$) also passed $t$-tests with $p < 0.05$, indicating that the model does not appear to contain redundant parameters. The regression equation takes the form of

$E_a = c_0 + \sum_{i=1}^{16} c_i x_i,$ (S1)

where $x_i$ are ordered by the sizes of standardized $c_i$ ($z[c_i] = \frac{c_i x_i}{z[x_i]}$; $z[x_i]$ is $z$-scored $x_i$). The regression model achieved $R^2_{\text{train}} = 0.786$, root mean squared error $\text{RMSE}_{\text{train}} = 182$ meV, and mean absolute error $\text{MAE}_{\text{train}} = 133$ meV for the training dataset, and $R^2_{\text{test}} = 0.664$ , $\text{RMSE}_{\text{test}} = 232$ meV, and $\text{MAE}_{\text{test}} = 163$ meV for the test dataset. When performing 100 random resamplings independently, each selecting 20 % of the total dataset as test set, the averaged performance metrics yield $\langle R^2 \rangle_{100} = 0.758$, $\langle \mathrm{RMSE} \rangle_{100} = 190$ meV, and $\langle \mathrm{MAE} \rangle_{100} = 138$ meV.

While the complete statistical details of $x_i$ $(i = 1, \cdots, 16)$ are provided in the section **“Full Description of $M_{f,1}$ for $E_a$”** in the **Supplementary Information**, I describe the most important terms $x_i$ $(i = 1, \cdots, 2)$ below:

$$x_1 = \sin\left(\frac{\pi}{10} E_1\right), \qquad \text{(S2)}$$

$$x_2 = \left(\sigma(\theta_D)\right)^2 E_1, \qquad \text{(S3)}$$

$$E_1 = \left[\operatorname{erf}\left(\frac{B_1}{A_1} - \frac{7}{8}\right)\right]^2, \qquad \text{(S4)}$$

$$A_1 = \exp\left(\operatorname{erf}\left(\left(\sigma(Z)\right)^2 \sin\left(\frac{\pi}{1000} \langle r_{\mathrm{VI}} \rangle \langle B \rangle\right)\right) - \frac{1}{2}\right) \operatorname{erf}\left(\bar{O} \langle \eta_f \rangle\right) \sin\left(\pi \frac{\bar{O}}{\langle r_{\mathrm{VI}} \rangle}\right), \qquad \text{(S5)}$$

and

$$B_1 = \sin\left(\pi \operatorname{erf}\left(\frac{\bar{O}}{\langle n \rangle} \sin\left(\frac{\pi}{100} \langle r_{\mathrm{VI}} \rangle \langle B \rangle\right) - \frac{1}{10}\right) \left[\exp\left(\left(\log_{10} \frac{\langle \rho_{\mathrm{mol}} \rangle}{\langle G \rangle}\right)^{-3}\right)\right]^{-2}\right), \qquad \text{(S6)}$$

Despite the algebraic complexity of **Eqs. (S2)—(S6)** their physical interpretation is clear as shown in **Supplementary Fig. 2b**. Considering that $\frac{B_1}{A_1} < \frac{7}{8}$ holds across all materials in the database, the following conditions effectively reduce $E_a$ inside $M_{f,1}$: (i) low charge disorder [$\sigma(Z)$, leading to positively small $A_1$], under which oxygen ions are less strongly bound to specific cations, thereby diminishing electrostatic constraints; (ii) a low oxygen ratio ($\bar{O}$, leading to positively small $A_1$), which facilitates greater ion mobility; (iii) loose packing, reflected by a small $\langle \rho_{\mathrm{mol}} \rangle$ (leading to negatively large $B_1$), which reduces steric hindrance to oxygen ion transport; and (iv) a high shear modulus ($\langle G \rangle$, leading to negatively large $B_1$), which helps preserve conduction pathways against emergent shear stresses generated during concerted oxygen ion diffusion within the cation framework; a low $\langle G \rangle$ serves as a penalizing factor for $E_a$.

Next, a prediction capability of $M_{f,1}$ for $A$ is shown in **Supplementary Fig. 2c**. The model satisfied the $F$-test with $p < 10^{-20}$, and $c_i$ and $c_0$ also passed $t$-tests with $p < 0.01$. The regression equation takes the form of

$$\log_{10} A = c_0 + \sum_{i=1}^{14} c_i x_i, \qquad \text{(S7)}$$

which achieved $R^2_{\text{train}} = 0.698$, $\text{RMSE}_{\text{train}} = 1.21\ \log_{10}[\text{K} \cdot \text{S} \cdot \text{cm}^{-1}]$, and $\text{MAE}_{\text{train}} = 0.867\ \log_{10}[\text{K} \cdot \text{S} \cdot \text{cm}^{-1}]$ for the training dataset, and $R^2_{\text{test}} = 0.582$, $\text{RMSE}_{\text{test}} = 1.37\ \log_{10}[\text{K} \cdot \text{S} \cdot \text{cm}^{-1}]$, and $\text{MAE}_{\text{test}} = 1.02\ \log_{10}[\text{K} \cdot \text{S} \cdot \text{cm}^{-1}]$ for the test dataset. The averaged performance metrics were $\langle R^2 \rangle_{100} = 0.678$, $\langle \text{RMSE} \rangle_{100} = 1.23\ \log_{10}[\text{K} \cdot \text{S} \cdot \text{cm}^{-1}]$, and $\langle \text{MAE} \rangle_{100} = 0.893\ \log_{10}[\text{K} \cdot \text{S} \cdot \text{cm}^{-1}]$.

While the complete statistical details of $x_i$ $(i = 1, \cdots, 14)$ are provided in the section **"Full Description of $M_{f,1}$ for $A$"** in the **Supplementary Information**, I describe the most important terms $x_i$ $(i = 1, \cdots, 2)$ below:

$$x_1 = \langle M \rangle, \qquad \text{(S8)}$$

$$x_2 = \sqrt[3]{\frac{B_2}{[\log_{10}(E_2 B_2)]^2} \frac{\langle M \rangle \langle \nu \rangle \langle \kappa \rangle}{\bar{O}}}, \quad \text{(S9)}$$

$$B_2 = \left[\text{erf}\left(\frac{\langle \nu \rangle \langle \theta_D \rangle}{100}\right)\right]^2, \qquad \text{(S10)}$$

and

$$E_2 = \left[\operatorname{erf}\left(\left[\operatorname{erf}\left(\frac{\langle \chi - \chi_{\mathrm{O}} \rangle}{10\sqrt{\langle r_{\mathrm{VI}} \rangle \langle \rho_{\mathrm{mol}} \rangle}}\right)\right]^2 \sin\left(\frac{\pi}{10}\frac{\langle \chi - \chi_{\mathrm{O}} \rangle}{A_2}\right) - \frac{1}{20}\right)\right]^2, \quad \text{(S11)}$$

Regarding the two most influential descriptors, $x_1$ and $x_2$, both are negatively correlated with $E_a$ ($c_1 < 0$ and $c_2 < 0$), meaning that their increase tends to reduce $A$. **Eq. (S8)** clearly shows that light metal atoms (small $\langle M \rangle$) are advantageous, as they increase the attempt frequency of ionic vibrations and thereby raise $A$ (**Supplementary Fig. 2d**). Despite the complexity of **Eqs. (S9)—(S11)**, the dominant contribution can be captured by

$$E_2^{\circ} = -\sqrt[3]{\frac{B_2}{[\log_{10}(E_2 B_2)]^2}} \qquad \text{(S12)}$$

which is positively correlated with $A$. As illustrated in **Supplementary Fig. 2e,** where the parameters $\langle r_{\mathrm{VI}} \rangle \langle \rho_{\mathrm{mol}} \rangle = 0.07$, $B_2 = 0.7$, and $\sigma(G) = 0$ were adopted for simplicity, high electronegativity ($\chi$, leading to the weak ionic bond) and high orbital anisotropy ($l$) drive high $E_2^{\circ}$, thereby enhancing $A$. From these results, three design principles emerge for maximizing the $A$ inside $M_{f,1}$: (i) low average atomic mass $(A)$ that favors higher vibrational attempt frequency, (ii) small electronegativity difference (or high $\langle \chi - \chi_{\mathrm{O}} \rangle$; note that $\langle \chi - \chi_{\mathrm{O}} \rangle < 0$) that leads to weaker ionic bonding, making oxygen ions less tightly bound to cations, (iii) high orbital anisotropy ($\langle l \rangle$) wherein $d$- and $f$-electrons may provide rich vibrational modes that couple effectively to ionic hopping.[S4, S5]

Additionally, **Supplementary Table 2** presents the performance metrics $R^2_{\mathrm{all}}$, $\mathrm{RMSE}_{\mathrm{all}}$ and $\mathrm{MAE}_{\mathrm{all}}$ for each individual model $M_{f,i_f}$ ($1 \le i_f \le N_f = 10$) constructed for $E_a$ and $A$. Most of the performance metrics exhibit consistent values across the models, with few anomalous or peculiar cases observed.

**Supplementary Table 2** Performance metrics $R^2_{\text{all}}$, $\text{RMSE}_{\text{all}}$ and $\text{MAE}_{\text{all}}$ for each individual model $M_{f,i_f}$ constructed for $E_a$ and $A$ with the number of included terms $n_t$. The units for $\text{RMSE}_{\text{all}}$ and $\text{MAE}_{\text{all}}$ are given in meV for $E_a$ and in $\log_{10}[\text{K} \cdot \text{S} \cdot \text{cm}^{-1}]$ for $\log_{10} A$, respectively.

| $E_a$ | $n_t$ | $R^2_{\text{all}}$ | $\text{RMSE}_{\text{all}}$ | $\text{MAE}_{\text{all}}$ | $A$ | $n_t$ | $R^2_{\text{all}}$ | $\text{RMSE}_{\text{all}}$ | $\text{MAE}_{\text{all}}$ |
|---|---|---|---|---|---|---|---|---|---|
| $M_{f,1}$ | 16 | 0.761 | 193 | 140 | $M_{f,1}$ | 14 | 0.677 | 1.24 | 0.899 |
| $M_{f,2}$ | 16 | 0.728 | 206 | 154 | $M_{f,2}$ | 12 | 0.661 | 1.27 | 0.909 |
| $M_{f,3}$ | 13 | 0.760 | 194 | 143 | $M_{f,3}$ | 18 | 0.671 | 1.25 | 0.906 |
| $M_{f,4}$ | 17 | 0.755 | 196 | 145 | $M_{f,4}$ | 12 | 0.692 | 1.21 | 0.899 |
| $M_{f,5}$ | 13 | 0.775 | 188 | 140 | $M_{f,5}$ | 16 | 0.685 | 1.23 | 0.884 |
| $M_{f,6}$ | 14 | 0.719 | 210 | 158 | $M_{f,6}$ | 10 | 0.638 | 1.31 | 0.971 |
| $M_{f,7}$ | 15 | 0.743 | 200 | 147 | $M_{f,7}$ | 18 | 0.569 | 1.44 | 1.07 |
| $M_{f,8}$ | 16 | 0.754 | 196 | 146 | $M_{f,8}$ | 20 | 0.644 | 1.30 | 0.941 |
| $M_{f,9}$ | 18 | 0.744 | 200 | 143 | $M_{f,9}$ | 18 | 0.658 | 1.28 | 0.921 |
| $M_{f,10}$ | 12 | 0.742 | 201 | 151 | $M_{f,10}$ | 14 | 0.624 | 1.21 | 0.880 |

## Full Description of $M_{f,1}$ for $E_a$

The full statistical details of descriptors $x_i$ ($i = 1, \cdots, 16$) are provided below:

$$x_1 = \sin\left(\frac{\pi}{10} E_1\right), \qquad \text{(S13)}$$

$$x_2 = \left(\sigma(\theta_D)\right)^2 E_1, \qquad \text{(S14)}$$

$$x_3 = [\log_{10}(\langle M \rangle \langle n_c \rangle)]^2, \qquad \text{(S15)}$$

$$x_4 = \frac{\langle B \rangle \langle n_c \rangle \langle \upsilon \rangle \langle \theta_D \rangle 10^{-\left[\mathrm{erf}\left(\langle Z \rangle \langle \eta_f \rangle\right) \mathrm{erf}\left(\sigma(r_{\mathrm{VI}}) - \frac{1}{20}\right)\right]^2}}{A_1}, \quad \text{(S16)}$$

$$x_5 = 10^{3\frac{\sqrt[3]{\langle \rho_{\mathrm{mol}} \rangle \langle G \rangle}}{\langle M \rangle \langle n_c \rangle}}, \qquad \text{(S17)}$$

$$x_6 = [\sigma(\theta_D)]^2, \qquad \text{(S18)}$$

$$x_7 = \exp(E_1 D_1), \qquad \text{(S19)}$$

$$x_8 = \sin\left(\frac{\langle \rho \rangle}{\langle \theta_D \rangle}\right), \qquad \text{(S20)}$$

$$x_9 = [\log_{10}([F_1 G_1]^2)]^{-3}, \qquad \text{(S21)}$$

$$x_{10} = \left[\exp\left({F_1}^2 \cos\left(10 \frac{\bar{O}}{\langle G \rangle}\right)\right)\right]^{-1}, \qquad \text{(S22)}$$

$$x_{11} = \cos\left(\pi E_1 \sqrt[3]{\langle \rho_{\mathrm{mol}} \rangle \langle G \rangle}\right), \quad \text{(S23)}$$

$$x_{12} = \left(\sigma(\theta_D)\right)^2 D_1, \qquad \text{(S24)}$$

$$x_{13} = \left[\mathrm{erf}\left(\frac{{D_1}^3 {G_1}^2}{A_1}\right)\right]^3, \quad \text{(S25)}$$

$$x_{14} = \frac{\langle \chi - \chi_{\mathrm{O}} \rangle}{\langle \rho \rangle}, \quad \text{(S26)}$$

$$x_{15} = \sin\left(\frac{\pi}{20} r(Z)\right), \quad \text{(S27)}$$

and

$$x_{16} = \cos\left(\frac{\pi}{10} \bar{O} \langle M \rangle\right). \quad \text{(S28)}$$

It is given that

$$A_1 = \exp\left(\mathrm{erf}\left(\left(\sigma(Z)\right)^2 \sin\left(\frac{\pi}{10} \langle r_{\mathrm{VI}} \rangle \langle B \rangle\right)\right) - \frac{1}{2}\right) \mathrm{erf}\left(\bar{O} \langle \eta_f \rangle\right) \sin\left(\pi \frac{\bar{O}}{\langle r_{\mathrm{VI}} \rangle}\right)\Bigg), \qquad \text{(S29)}$$

$$B_1 = \sin\left(\pi\, \mathrm{erf}\left(\frac{\bar{O}}{\langle n \rangle} \sin\left(\frac{\pi}{1000} \langle r_{\mathrm{VI}} \rangle \langle B \rangle\right) - \frac{1}{10}\right) \left[\exp\left(\left(\log_{10} \frac{\langle \rho_{\mathrm{mo}} \rangle}{\langle G \rangle}\right)^{-3}\right)\right]^{-2}\right), \qquad \text{(S30)}$$

$$C_1 = \left[\sin\left(\frac{\pi}{10} \frac{\langle \rho_{\mathrm{mol}} \rangle}{\langle \eta_f \rangle}\right)\right] \left[\exp\left(\sin\left(\frac{1}{200\pi} r(l)\right) \cos\left(\frac{\pi}{10} r(\kappa)\right)\right)\right]^{-2}, \qquad \text{(S31)}$$

$$D_1 = \mathrm{erf}\left(\left(\sigma(Z)\right)^2 \sin\left(\frac{\pi}{1000} \langle r_{\mathrm{VI}} \rangle \langle B \rangle\right) - \frac{1}{2}\right) \sin\left(\frac{\pi}{10} \frac{\langle \rho_{\mathrm{mo}} \rangle}{\langle \eta_f \rangle}\right) 10^{\frac{\bar{O}}{\langle n \rangle} \cos\left(\frac{\pi}{10} r(\kappa)\right) 10^{\frac{\bar{O}}{\langle n \rangle} \sin\left(\frac{\pi}{100} \langle \chi - \chi_{\mathrm{O}} \rangle \langle \rho \rangle\right)}}, \qquad \text{(S32)}$$

$$E_1 = \left[\mathrm{erf}\left(\frac{B_1}{A_1} - \frac{7}{8}\right)\right]^2, \quad \text{(S33)}$$

$$F_1 = \mathrm{erf}\left(\frac{C_1}{A_1} - \frac{7}{8}\right), \qquad \text{(S34)}$$

and

$$G_1 = \mathrm{erf}\left(\frac{\bar{O}}{\langle Z \rangle} - 1\right). \qquad \text{(S35)}$$

I leave the statistical details in **Supplementary Table 3** and represent residual histogram in **Supplementary Fig. 3.** The latter showing zero-centered distributions, this result demonstrates that the model errors are random rather than systematic, with no apparent bias or pattern.

**Supplementary Table 3** Coefficient ($c_i$), standardized coefficient ($z[c_i]$), standard error ($SE$), *t*-test value, and 95%-confidential intervals ($[C_1, C_2]$) for each descriptor $x_i$ and the intercept term ($c_0$) of $E_a$.

| $x_i$ | $c_i$ | $z[c_i]$ | $SE$ | $t$ | $C_1$ | $C_2$ |
|---|---|---|---|---|---|---|
| $x_1$ | -19100 | -2.29 | 1200 | -16.0 | -21700 | -16500 |
| $x_2$ | -0.0531 | -1.61 | 0.0077 | -6.89 | -0.0698 | -0.0364 |
| $x_3$ | 613 | 1.53 | 45.7 | 13.4 | 514 | 712 |
| $x_4$ | -0.0253 | -1.50 | 0.00144 | -17.6 | -0.0284 | -0.0222 |
| $x_5$ | 95400 | 1.08 | 7600 | 12.6 | 78900 | 112000 |
| $x_6$ | 0.0229 | 1.00 | 0.00546 | 4.19 | 0.0111 | 0.0347 |
| $x_7$ | -4290 | -0.972 | 368 | -11.6 | -5080 | -3490 |
| $x_8$ | -20500 | -0.895 | 2130 | -9.61 | -25100 | -15900 |
| $x_9$ | -1050 | -0.740 | 76.3 | -13.8 | -1220 | -885 |
| $x_{10}$ | 1230 | 0.518 | 168 | 7.30 | 862 | 1590 |
| $x_{11}$ | -579 | -0.484 | 105 | -5.49 | -807 | -350 |
| $x_{12}$ | -0.0300 | -0.405 | 0.00611 | -4.92 | -0.0433 | -0.0168 |
| $x_{13}$ | 28900 | 0.397 | 2370 | 12.2 | 23800 | 34100 |
| $x_{14}$ | -980 | -0.360 | 213 | -4.61 | -1440 | -519 |
| $x_{15}$ | 224 | 0.132 | 48.8 | 4.58 | 118 | 329 |
| $x_{16}$ | -34.2 | -0.0603 | 14.6 | -2.34 | -65.9 | -2.47 |
| $c_0$ | -93800 | 0 | 8120 | -11.6 | -111000 | -76200 |

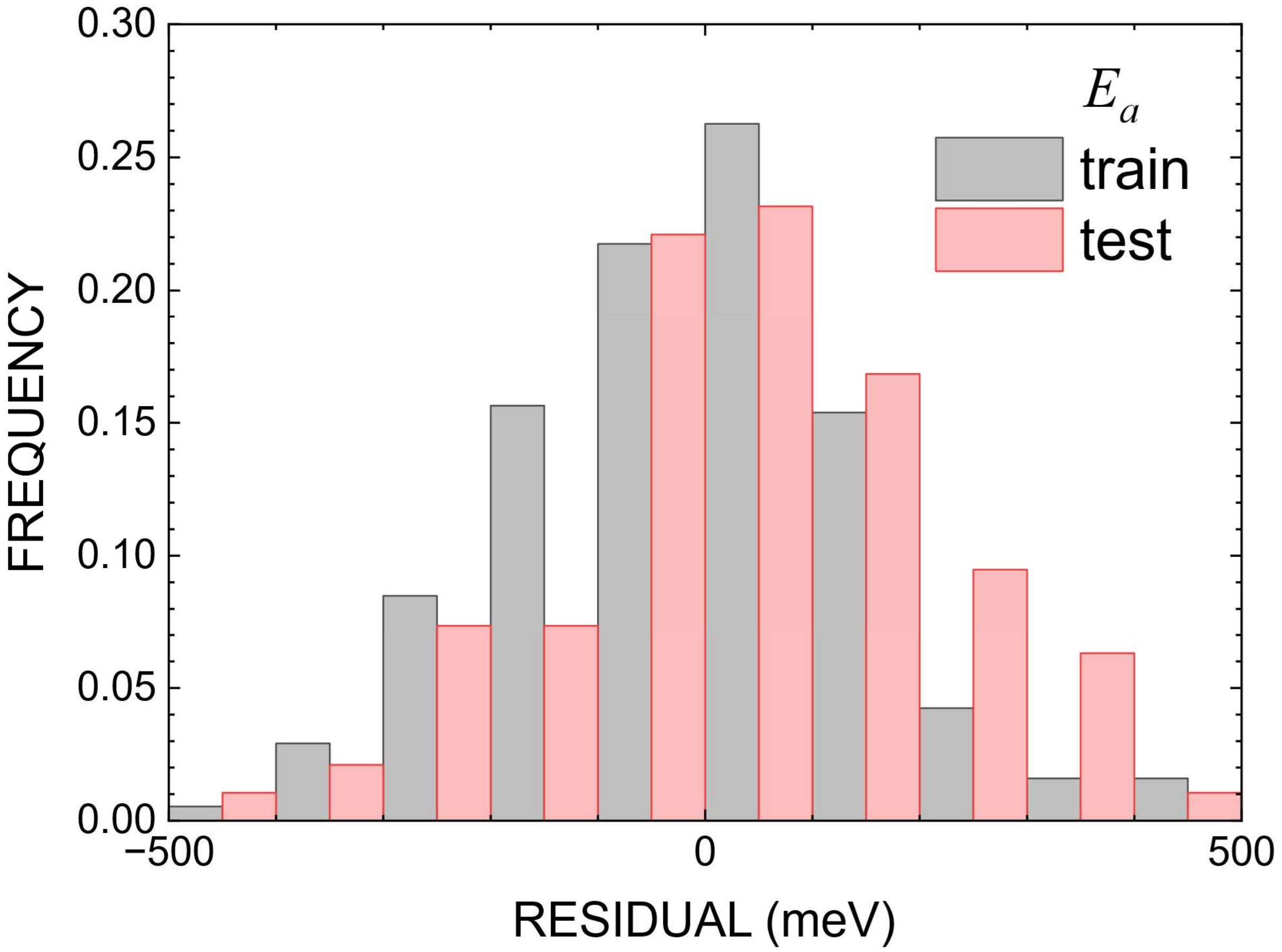

**Supplementary Fig. 3** Residual histogram plot for the regression model for $E_a$. The training and test datasets are represented by gray and red blocks, respectively.

**Full Description of $M_{f,1}$ for $A$**

The full statistical details of descriptors $x_i$ ($i = 1, \cdots, 14$) are provided below:

$$x_1 = \langle M \rangle, \qquad \text{(S36)}$$

$$x_2 = \sqrt[3]{\frac{B_2}{[\log_{10}(E_2 B_2)]^2} \frac{\langle M \rangle \langle \upsilon \rangle \langle \kappa \rangle}{\bar{O}}}, \quad \text{(S37)}$$

$$x_3 = \left[\mathrm{erf}\left(\frac{\langle r_{\mathrm{VI}} \rangle \langle \rho \rangle}{10}\right)\right]^2, \quad \text{(S38)}$$

$$x_4 = \sin\left(\frac{1}{100 C_2 F_2 B_2 \langle M \rangle \langle \upsilon \rangle}\right), \qquad \text{(S39)}$$

$$x_5 = D_2 \sin\left(\frac{\pi}{1000} \sin\left(\frac{\pi}{1000} \sigma(B)\right) E_2 \langle G \rangle \langle \theta_D \rangle \langle \chi - \chi_{\mathrm{O}} \rangle [\mathrm{erf}(\sigma(\chi - \chi_{\mathrm{O}}) - 1)]^2\right), \qquad \text{(S40)}$$

$$x_6 = \left[\mathrm{erf}\left(\mathrm{erf}(\sigma(Z) - 1) \left(\frac{\langle \rho_{\mathrm{mol}} \rangle B_2}{\langle n \rangle}\right)^2 \langle M \rangle \langle \nu \rangle - \frac{1}{20}\right)\right]^2, \qquad \text{(S41)}$$

$$x_7 = \mathrm{erf}\left(G_2 \, \mathrm{erf}\left(D_2 \left[\mathrm{erf}\left(\frac{E_2}{B_2}\right)\right]^2 - \frac{1}{2}\right) - \frac{7}{8}\right), \quad \text{(S42)}$$

$$x_8 = [\langle B \rangle \langle n_c \rangle]^3 \cos\left(\frac{\langle M \rangle}{10 \langle B \rangle}\right), \quad \text{(S43)}$$

$$x_9 = \cos\left(\frac{\pi}{2000} \frac{F_2 B_2}{C_2} \langle M \rangle \langle \upsilon \rangle\right), \quad \text{(S44)}$$

$$x_{10} = \langle \upsilon \rangle \langle r_{\mathrm{VI}} \rangle, \qquad \text{(S45)}$$

$$x_{11} = [\mathrm{erf}(\sigma(\chi - \chi_{\mathrm{O}}) - 1)]^2, \qquad \text{(S46)}$$

$$x_{12} = E_2 G_2 \frac{\langle r_{\mathrm{VI}} \rangle}{\langle \rho \rangle} \left[\frac{\langle B \rangle \langle r \rangle \langle \rho \rangle}{\langle \upsilon \rangle [\exp(\langle \rho_{\mathrm{mol}} \rangle \langle n_{\mathrm{c}} \rangle)]^3} \sin(\pi \sigma(n)) \cos\left(\frac{\pi}{100} A_2 [\langle G \rangle \langle \theta_{\mathrm{D}} \rangle]^2 \left[\log_{10}\left(\frac{\bar{O}}{\langle G \rangle}\right)\right]^6\right)\right]^3, \quad \text{(S47)}$$

$x_{13} = \left[\log_{10}\left(\frac{\langle M \rangle}{\langle B \rangle}\right)\right]^2$, (S48)

and

$x_{14} = \mathrm{erf}\big(r(\rho_{\mathrm{mol}})\big)$. (S49)

It is given that

$$A_2 = 10^{-3 \sin\left(\frac{\pi}{100}\langle \chi - \chi_{\mathrm{O}} \rangle \langle l \rangle\right) \sin\left(\frac{\pi}{200}\sigma(G)\right)}, \qquad \text{(S50)}$$

$$B_2 = \left[\mathrm{erf}\left(\frac{\langle v \rangle \langle \theta_D \rangle}{100}\right)\right]^2, \text{(S51)}$$

$$C_2 = \left[\exp\left(\left[\mathrm{erf}\left(10\left[\mathrm{erf}\left(\sqrt[3]{\langle M \rangle \langle \kappa \rangle}\sqrt{\sigma(\nu)} - \frac{1}{20}\right)\right]^2 \langle M \rangle \langle \nu \rangle\right)\right]^2\right)\right]^2, \text{(S52)}$$

$$D_2 = \exp\left(\left[\mathrm{erf}\left([\log_{10}(\langle l \rangle \langle n_c \rangle)]^2 \sin\left(100\frac{\langle \eta_f \rangle}{\langle n \rangle}\right) - 1\right)\right]^2\right), \text{(S53)}$$

$$E_2 = \left[\mathrm{erf}\left(\left[\mathrm{erf}\left(\frac{\langle \chi - \chi_{\mathrm{O}} \rangle}{10\sqrt{\langle r_{\mathrm{VI}} \rangle \langle \rho_{\mathrm{mol}} \rangle}}\right)\right]^2 \sin\left(\frac{\pi}{10}\frac{\langle \chi - \chi_{\mathrm{O}} \rangle}{A_2}\right) - \frac{1}{20}\right)\right]^2, \text{(S54)}$$

$$F_2 = \left[\mathrm{erf}\left(\frac{1}{1000}\sqrt{A_2\left[\log_{10}\left(\frac{\bar{O}}{\langle G \rangle}\right)\right]^{-3}\frac{\langle \chi - \chi_{\mathrm{O}} \rangle}{\langle \kappa \rangle}}\right)\right]^2, \qquad \text{(S55)}$$

and

$$G_2 = \sqrt[3]{\frac{\langle \rho_{\mathrm{mol}} \rangle \langle \kappa \rangle}{F_2}}. \qquad \text{(S56)}$$

I leave the statistical details in **Supplementary Table 4** and represent residual histogram in **Supplementary Fig. 4.** The latter showing zero-centered distributions, this result demonstrates that the model errors are random rather than systematic, with no apparent bias or pattern.

**Supplementary Table 4** Coefficient ($c_i$), standardized coefficient ($z[c_i]$), standard error ($SE$), *t*-test value, and 95%-confidential intervals ($[C_1, C_2]$) for each descriptor $x_i$ and the intercept term ($c_0$) of $\log_{10} A$.

| $x_i$ | $c_i$ | $z[c_i]$ | $SE$ | $t$ | $C_1$ | $C_2$ |
|---|---|---|---|---|---|---|
| $x_1$ | -0.0427 | -0.782 | 0.00807 | -5.29 | -0.0601 | -0.0252 |
| $x_2$ | -113 | -0.725 | 9.75 | -11.6 | -134 | -92.1 |
| $x_3$ | 7.67 | 0.701 | 1.48 | 5.17 | 4.46 | 10.9 |
| $x_4$ | 493 | 0.640 | 34.4 | 14.3 | 418 | 567 |
| $x_5$ | 7.92 | 0.536 | 0.616 | 12.9 | 6.58 | 9.25 |
| $x_6$ | 2050 | 0.391 | 202 | 10.1 | 1610 | 2490 |
| $x_7$ | -2.09 | -0.317 | 0.274 | -7.64 | -2.69 | -1.50 |
| $x_8$ | $-1.52\times10^{-9}$ | -0.309 | $1.96\times10^{-10}$ | -7.77 | $-1.94\times10^{-9}$ | $-1.10\times10^{-9}$ |
| $x_9$ | -98700 | -0.283 | 14500 | -6.82 | -130000 | -67300 |
| $x_{10}$ | 8.22 | 0.213 | 1.42 | 5.78 | 5.14 | 11.3 |
| $x_{11}$ | -3.49 | -0.193 | 0.645 | -5.41 | -4.89 | -2.09 |
| $x_{12}$ | $-6.01\times10^{-8}$ | -0.178 | $9.76\times10^{-9}$ | -6.15 | $-8.12\times10^{-8}$ | $-3.89\times10^{-8}$ |
| $x_{13}$ | 1.54 | 0.159 | 0.485 | 3.18 | 0.49 | 2.60 |
| $x_{14}$ | 0.504 | 0.118 | 0.182 | 2.77 | 0.110 | 0.898 |
| $c_0$ | 98700 | 0 | 14500 | 6.82 | 67300 | 130000 |

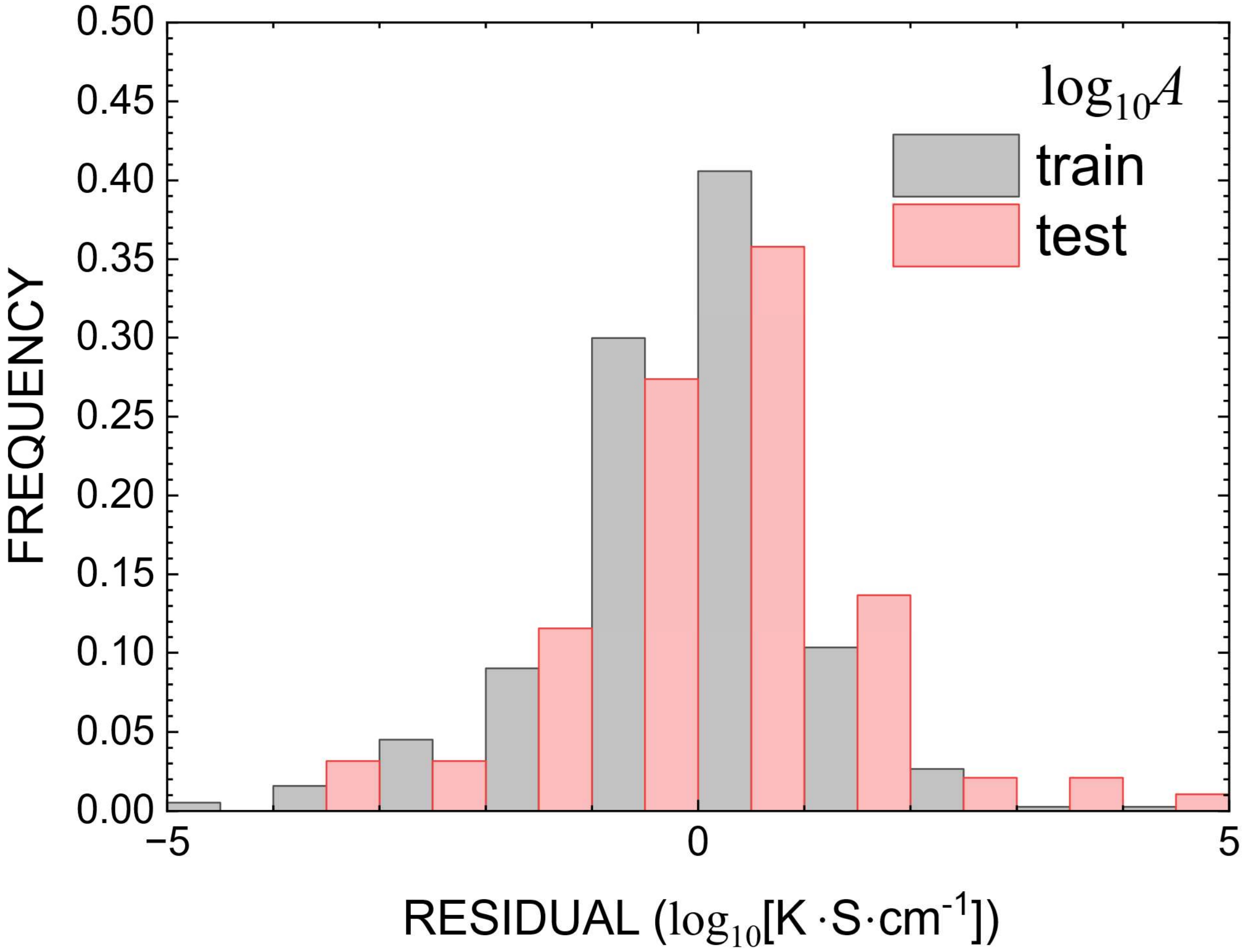

**Supplementary Fig. 4** Residual histogram plot for the regression model for $\log_{10} A$. The training and test datasets are represented by gray and red blocks, respectively.

**Important Interaction Search**

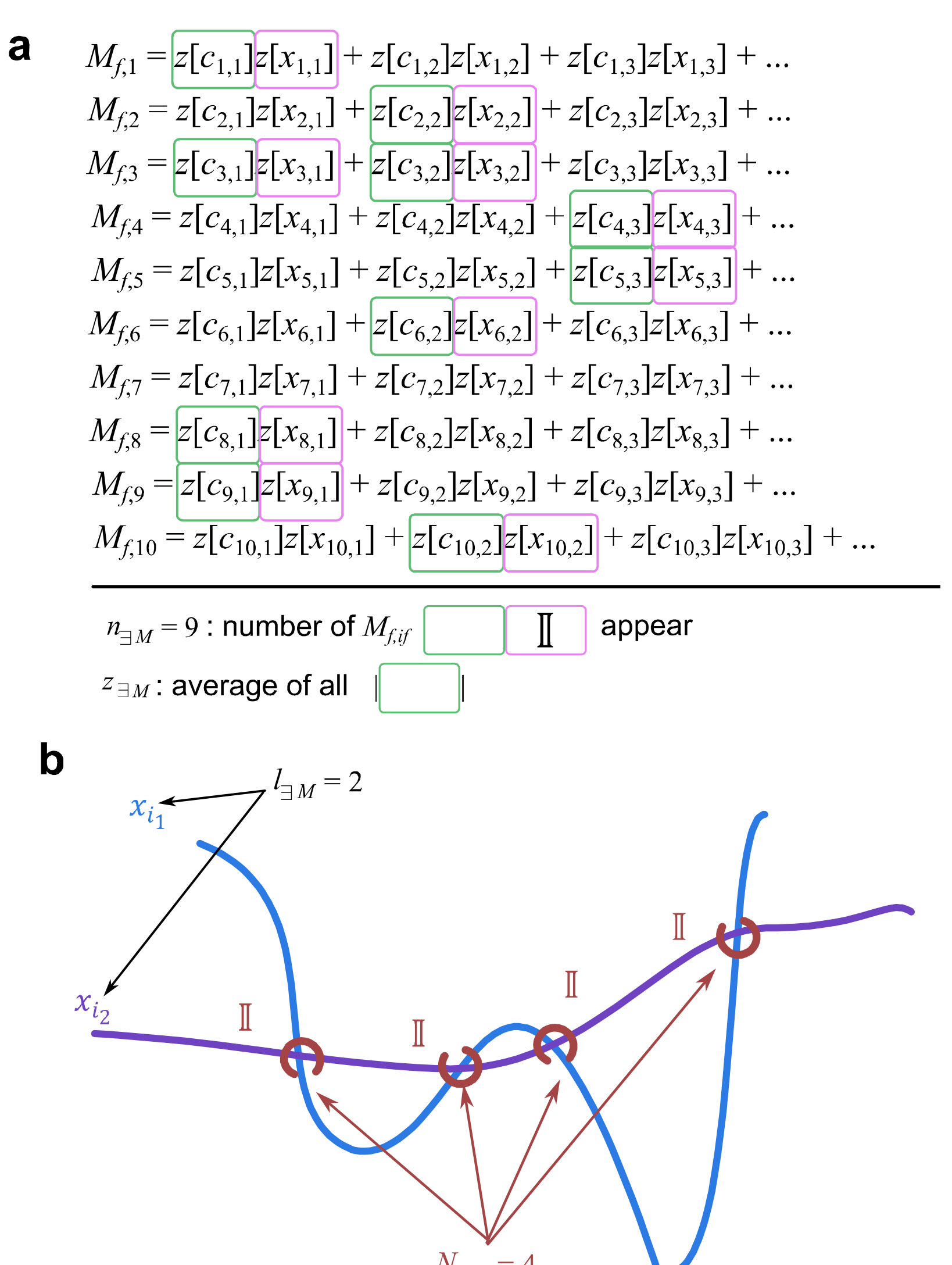

**Supplementary Fig. 5** Identification of important interaction sets $\mathbb{I}$ and chains $\mathbb{I}^*$. (a) Schematic illustration of frequency of appearance $n_{\exists M}$ and weighted-average coefficient magnitude $z_{\exists M}$ for an interaction set $\mathbb{I}$, which is an element of $\mathbb{I}^*$. (b) Analogous to concepts in knot theory: larger $z_{\exists M}$ values indicate persistent interactions among $l_{\exists M}$ strands with crossing numbers $n_{\exists M}$, reflecting their entanglement.

One of the advantages of employing symbolic regression models lies in their ability to reveal not only the important individual features but also the key interactions among features. Such interactions are explicitly represented through the multiplication or division of (scalar-transformed) features $x_{i_i}$ that co-occur within a term $x_{i_f,i}$ in $M_{f,i_f}$, written as

$$M_{f,i_f} = c_{i_f,0} + \sum_i c_{i_f,i} x_{i_f,i} = \mathrm{z}\left[c_{i_f,i}\right] \mathrm{z}\left[x_{i_f,i}\right] + \cdots . \text{(S57)}$$

To identify these interactions, a set of interacting features is defined:

$$\mathbb{I} = \{x_{i_i} \mid x_{i_i} \in X_1^{\circ}\} \quad \text{(S58)}$$

and refer to its interaction level as

$$l_{\exists M}(\mathbb{I}) = n(\mathbb{I}), \qquad \text{(S59)}$$

which corresponds to the number of features jointly appearing in $\mathbb{I}$. Then, two quantitative measures, $n_{\exists M}$ and $z_{\exists M}$, are introduced to evaluate the importance of such interaction sets, as illustrated in **Supplementary Fig. 5a**

**a. Frequency of appearance $n_{\exists M}$**

The first measure, $n_{\exists M}$, counts how many iteration models $M_{f,i_f}$ contain the interaction(s) $\mathbb{I}$. To formalize this, an "existence" function $\delta_{\exists x_{i_i}}$ is defined, which returns 0 if a term does not include $x_{i_i}$, and 1 otherwise. For a given interaction set $\mathbb{I}$, the joint existence of all its members is expressed as

$$\prod_{x_{i_i} \in I} \delta_{\exists x_{i_i}} . \quad \text{(S60).}$$

In the term vector of a model $M_{f,i_f}$, the set of terms can be represented as

$$\mathbb{T}_{M_{f,i_f}} = c_{i_f,i}x_{i_f,i}, \ldots \; = \mathrm{z}\left[c_{i_f,i}\right]\mathrm{z}\left[x_{i_f,i}\right], \ldots \, . \quad \text{(S61)}$$

Then, the number of models containing the interaction $\mathbb{I}$ is given by

$$n_{\exists M}(\mathbb{I} = \{x_{i_i} \mid x_{i_i} \in X_1^{\circ}\}) = \sum_{M_{f,i_f}} \left[1 - \delta\left(\sum\left(\prod_{x_{i_i} \in I} \delta_{\exists x_{i_i}}\right)\mathbb{T}_{M_{f,i_f}}\right)\right]. \qquad \text{(S62)}$$

For example, the interaction set $\mathbb{I} = \langle n_c \rangle, \bar{O}$ for $E_a$ represents a level-two interaction ($l_{\exists M}(\mathbb{I}) = 2$). Here, $n_{\exists M}$ counts how many models $M_{f,i_f}$ have terms where both $\langle n_c \rangle$ and $\bar{O}$ coexist. A representative case can be found in $x_4$ of $M_{f,1}$ , given by **Eqs. (S16) and (S29).**

**b. Weighted-average coefficient magnitude $z_{\exists M}$**

The second measure, $z_{\exists M}$, quantifies the average absolute magnitude of the coefficients associated with the interaction $\mathbb{I}$ across all models $M_{f,i_f}$ , weighted by $w_{i_f} = m_{i_f} / \sum_{k_f=1}^{N_f} \left| m_{k_f} \right|$ . It is algebraically defined as

$$z_{\exists M}(\mathbb{I} = \{x_{i_i} \mid x_{i_i} \in X_1^{\circ}\}) = \sum_{M_{f,i_f}} \left[\left(\prod_{x_{i_i} \in I} \delta_{\exists x_{i_i}}\right)\mathbb{T}_{M_{f,i_f}}\right]\left[\mathbb{Z}_{M_{f,i_f}}\right]^{T}, \text{(S63)}$$

where it is given that $\mathbb{Z}_{M_{f,i_f}} = \left\{w_{i_f}\left|z\left[c_{i_f,i}\right]\right|, \cdots\right\}$. For example, for $\mathbb{I} = \langle n_c \rangle, \bar{O}$ in $E_a$, all terms across all $M_{f,i_f}$ in which $\langle n_c \rangle$ and $\bar{O}$ coexist are collected, and compute the weighted-average of all corresponding $w_{i_f}|z[c_i]|$. Generally, a large $n_{\exists M}$ and a large $z_{\exists M}$ together indicate that the interaction $\mathbb{I}$ plays an important physical role.

It should be remarked that, conceptually, the important interaction sets $\mathbb{I}$ across ensemble models can be interpreted as a topological structure akin to a *knot or link diagram*, where each feature represents a strand and each interaction represents a crossing (see **Supplementary Fig. 5b**). Given the crossing numbers $n_{\exists M}$ of $l_{\exists M}$ strands, the persistence of certain interactions, represented by large $z_{\exists M}$, thus reflects invariant-like quantities describing the degree of entanglement among physical descriptors across diverse hermeneutical conversations $M_{f,i_f}$. This suggests a potential connection between symbolic regression and knot-theoretical representations of complex relationships.

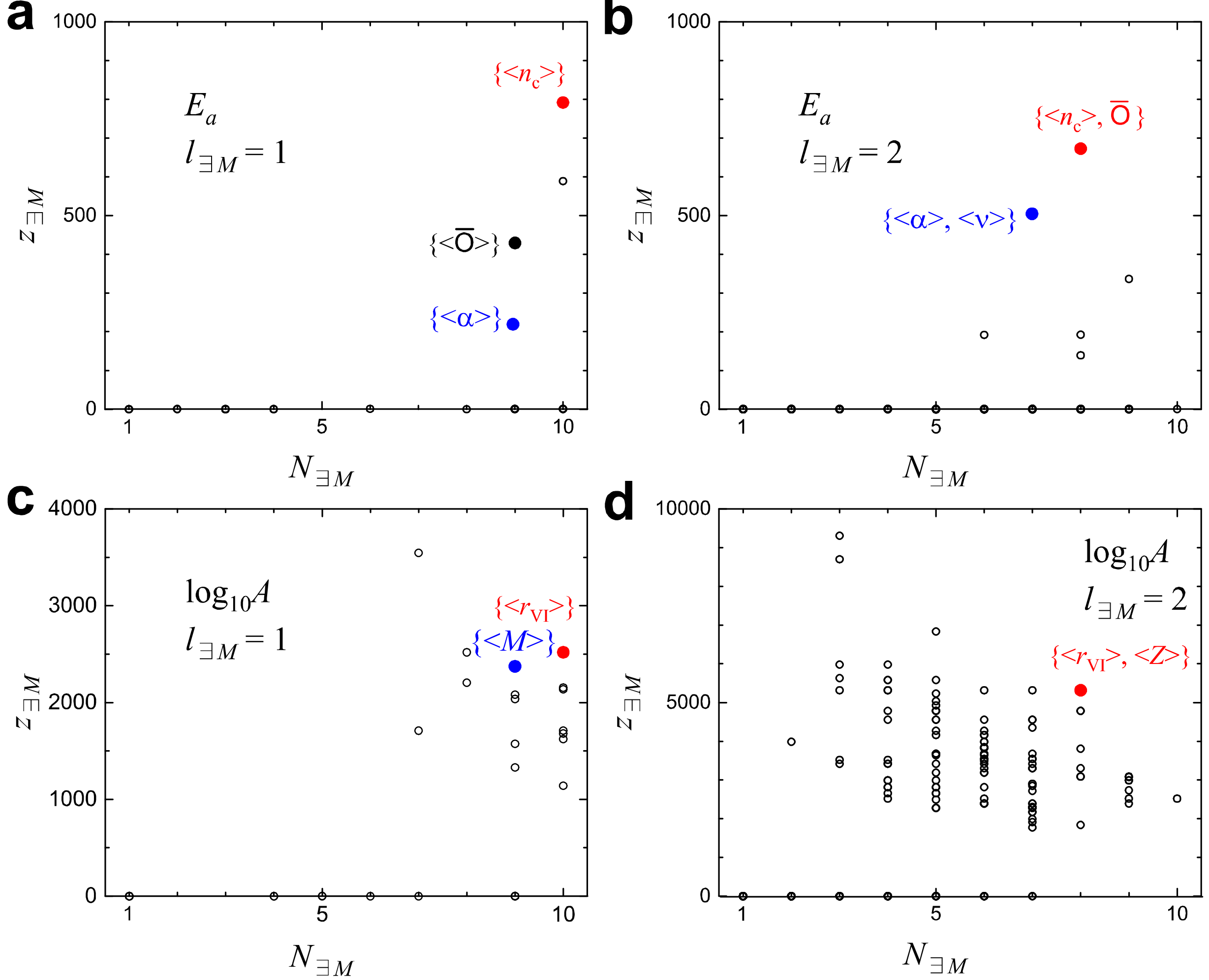

**Supplementary Fig. 6** Important interactions $\mathbb{I}$ for $E_a$ and $\log_{10} A$. (a) The (second) most interaction level $l_{\exists M} = 1$ for $E_a$, represented by red (blue) dot: $\mathbb{I}|_{l_{\exists M}=1} = \{\langle n_c \rangle\}$ ($\{\langle \alpha \rangle\}$). (b) $l_{\exists M} = 2$ for $E_a$: $\mathbb{I}|_{l_{\exists M}=2} = \{\langle n_c \rangle, \bar{O}\}$ ($\{\langle \alpha \rangle, \langle \nu \rangle\}$). (c) $l_{\exists M} = 1$ for $\log_{10} A$: $\mathbb{I}|_{l_{\exists M}=1} = \{\langle r_{\mathrm{VI}} \rangle\}$ ($\{\langle M \rangle\}$). (d) $l_{\exists M} = 2$ for $\log_{10} A$: $\mathbb{I}|_{l_{\exists M}=2} = \{\langle r_{\mathrm{VI}} \rangle, \langle Z \rangle\}$. The horizontal $n_{\exists M}$ and vertical $z_{\exists M}$ axes represent the number of $M_{f,i_f}$ where the interaction(s) of $\mathbb{I}$ appear and the weighted-average size of $w_{i_f}|z[c_i]|$ for $\mathbb{I}$ across all $M_{f,i_f}$, respectively.

By measuring frequency of appearance $n_{\exists M}$ and weighted-average coefficient magnitude $z_{\exists M}$ for a given interaction with $l_{\exists M}$, it becomes possible to construct the most important feature chain $\mathbb{I}^* = \{\mathbb{I}\}$, reminiscent of Jean Cavaillès's concept of *concatenation*.

**a. At $l_{\exists M} = 1$:**

Select the feature set $\mathbb{I}|_{l_{\exists M}=1}$ with the highest $z_{\exists M}$ among those satisfying the largest $n_{\exists M}$. For example, for $E_a$, the feature $\mathbb{I}|_{l_{\exists M}=1} = \langle n_c \rangle$ appears in all models ($n_{\exists M} = 10$) and has the largest $z_{\exists M}$ among them (see the red dot in **Supplementary Fig. 6a**). Insert $\mathbb{I}|_{l_{\exists M}=1}$ into $\mathbb{I}^*$: $\mathbb{I}^* = \{\mathbb{I}|_{l_{\exists M}=1}\} = \{\langle n_c \rangle\}$.

**b. At $l_{\exists M} = 2$:**

Select the interaction $\mathbb{I}|_{l_{\exists M}=2}$ with the highest $z_{\exists M}$ among those satisfying $n_{\exists M} \geq 8$, i.e., reducing the minimum $n_{\exists M}$ criterion by two (from 10 that is the $n_{\exists M}$ value at $l_{\exists M} = 1$), which is the "superset" of $\mathbb{I}|_{l_{\exists M}=1}$. For instance, $\mathbb{I}|_{l_{\exists M}=2} = \langle n_c \rangle, \bar{O} \supset \mathbb{I}|_{l_{\exists M}=1}$ (see the red dot in **Supplementary Fig. 6b**). Insert $\mathbb{I}|_{l_{\exists M}=2}$ into $\mathbb{I}^*$: $\mathbb{I}^* = \{\mathbb{I}|_{l_{\exists M}=1}, \mathbb{I}|_{l_{\exists M}=2}\} = \{\langle n_c \rangle, \langle n_c \rangle, \bar{O}\}$. Higher-order interactions ($l_{\exists M} = 3,4, \ldots$) can be similarly identified, but for simplicity, the analysis is limited to $l_{\exists M} \leq 2$.

For $\log_{10} A$, the same procedure reveals that at $l_{\exists M} = 1$, the dominant feature is $\mathbb{I}|_{l_{\exists M}=1} = \langle r_{VI} \rangle$ (see the red dot in **Supplementary Fig. 6c**), and at $l_{\exists M} = 2$, the key interaction is $\mathbb{I}|_{l_{\exists M}=2} = \langle r_{VI} \rangle, \langle Z \rangle$ (see the red dot in **Supplementary Fig. 6d**): $\mathbb{I}^* = \{\mathbb{I}|_{l_{\exists M}=1}, \mathbb{I}|_{l_{\exists M}=2}\} = \{\langle r_{VI} \rangle, \langle r_{VI} \rangle, \langle Z \rangle\}$.

The program is also capable of adding chains starting with the second, third, fourth, ⋯ important feature(s) at $l_{\exists M} = 1$, by starting with the second, third, fourth, ⋯ largest $n_{\exists M}$, which will yield multiple $l_{\exists M}$-way partial dependence plots by fixing the other features at their average values across all data points. For example, starting with $n_{\exists M} = 9$, the second most important interaction chains $\mathbb{I}^*$ are given as $\mathbb{I}^* = \{\{\langle\alpha\rangle\}, \{\langle\alpha\rangle, \langle\nu\rangle\}\}$ for $E_a$ and $\mathbb{I}^* = \{\{\langle M\rangle\}\}$ for $\log_{10} A$ (see the blue dots in **Supplementary Figs. 6a-6d**). The set $\mathbb{I}|_{l_{\exists M}=2}$, constituting the superset of $\mathbb{I}|_{l_{\exists M}=1} = \{\langle M\rangle\}$, could not be retrieved. This absence suggests that $\langle M\rangle$ functions as an independent feature without detectable interactions with the remaining descriptors.

The multiple $l_{\exists M}$-way partial dependence plots clearly illustrate how each important interaction chain $\mathbb{I}^*$ specifically influences the target variable, providing simple but deeper physical insight into the underlying mechanisms. In Ref. S1, the most and second most important interaction chains $\mathbb{I}^*$ for $l_{\exists M} = 2$ were visualized using two-way partial dependence plots, together with their physical interpretations.

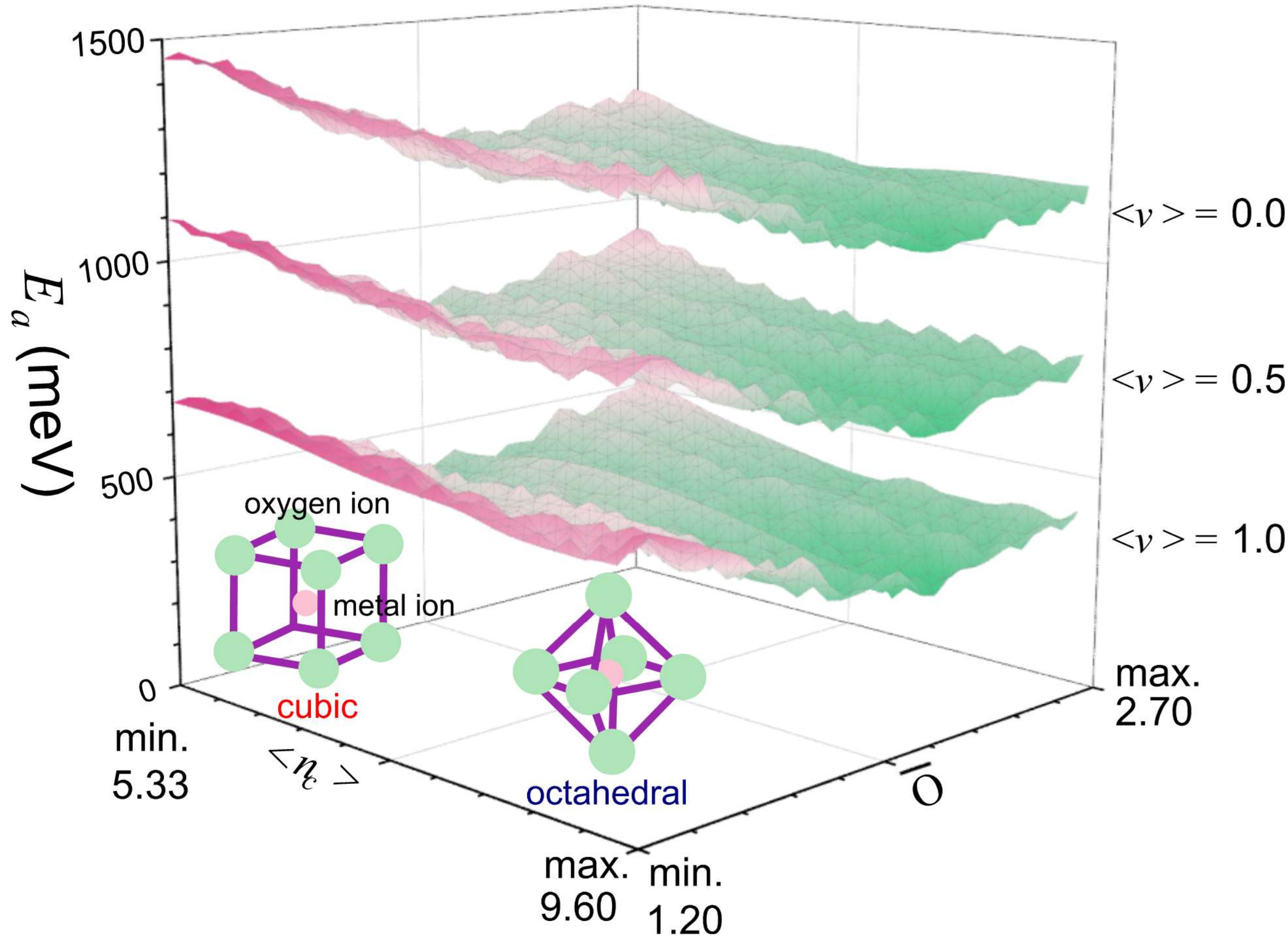

**Supplementary Fig. 7** Three-way partial dependence plot for $E_a$ showing the interaction among the three key features $\langle n_c \rangle$, $\bar{O}$, and $\langle \nu \rangle$. The minimum and maximum values of each feature are indicated. Each $\langle n_c \rangle$–$\bar{O}$ plane exhibits the similar landscapes of $E_a$, stratified according to $\langle \nu \rangle$. As illustrative examples, octahedral and cubic coordination cages of oxygen ions surrounding a metal ion are shown along the $\langle n_c \rangle$ axis.

**The designer module is also capable of generating high-$l_{\exists M}$-way partial dependence plots, which represent higher-order interactions among features. Supplementary Fig. 7** presents a three-way partial dependence plot for $E_a$, illustrating the capability of the designer module to elucidate underlying physical insights even for complex phenomena. It is noteworthy that, while the significant interaction at $l_{\exists M} = 3$ was identified as $\mathbb{I}|_{l_{\exists M}=3} = \langle n_c \rangle, \bar{O}, \langle \nu \rangle$, no comparably high-$l_{\exists M}$ interaction was observed for $\log_{10} A$. In the $\langle n_c \rangle$-$\bar{O}$ space, $E_a$ is minimized as both $\langle n_c \rangle$ and $\bar{O}$ increase. This trend suggests that oxygen-ion conduction is facilitated in environments with more densely packed surrounding oxygen ions, where electrostatic repulsion can help flatten the potential-energy landscape and promote ion migration. Consequently, $E_a$ is better described by the collective dynamics of multiple migrating ions rather than by the single-particle picture assumed in the nudged elastic band method. Moreover, the level of $E_a$ across the $\langle n_c \rangle$-$\bar{O}$ plane is strongly influenced by the Poisson's ratio $\langle \nu \rangle$. A larger $\langle \nu \rangle$ may imply that the lattice structure can "breathe" more easily, thereby facilitating ionic migration and lowering the activation barrier for oxygen ion conduction.

In addition, the designer module can export the model equations in copy-and-paste-ready LaTeX and Mathematica formats, thereby enabling users to perform further symbolic analyses. For example, the algebraically complex $M_{f,1}$ expression for $E_a$ can be simplified by fixing the other independent variables to their average values:

$$M_{f,1}(\langle n_c \rangle) = 95400\, \mathrm{e}^{\frac{0.0670}{\langle n_c \rangle}} - 125\, \langle n_c \rangle + 116\, (\log_e \langle n_c \rangle + 4.79)^2 - 99600. \qquad \text{(S64)}$$

This enables local interpretability at the level of individual features: information that is often hidden in the high-dimensional parameter space of conventional models, especially when describing complex physical phenomena.

**Reproducibility and Promising Materials for Oxygen-ion Conductors**

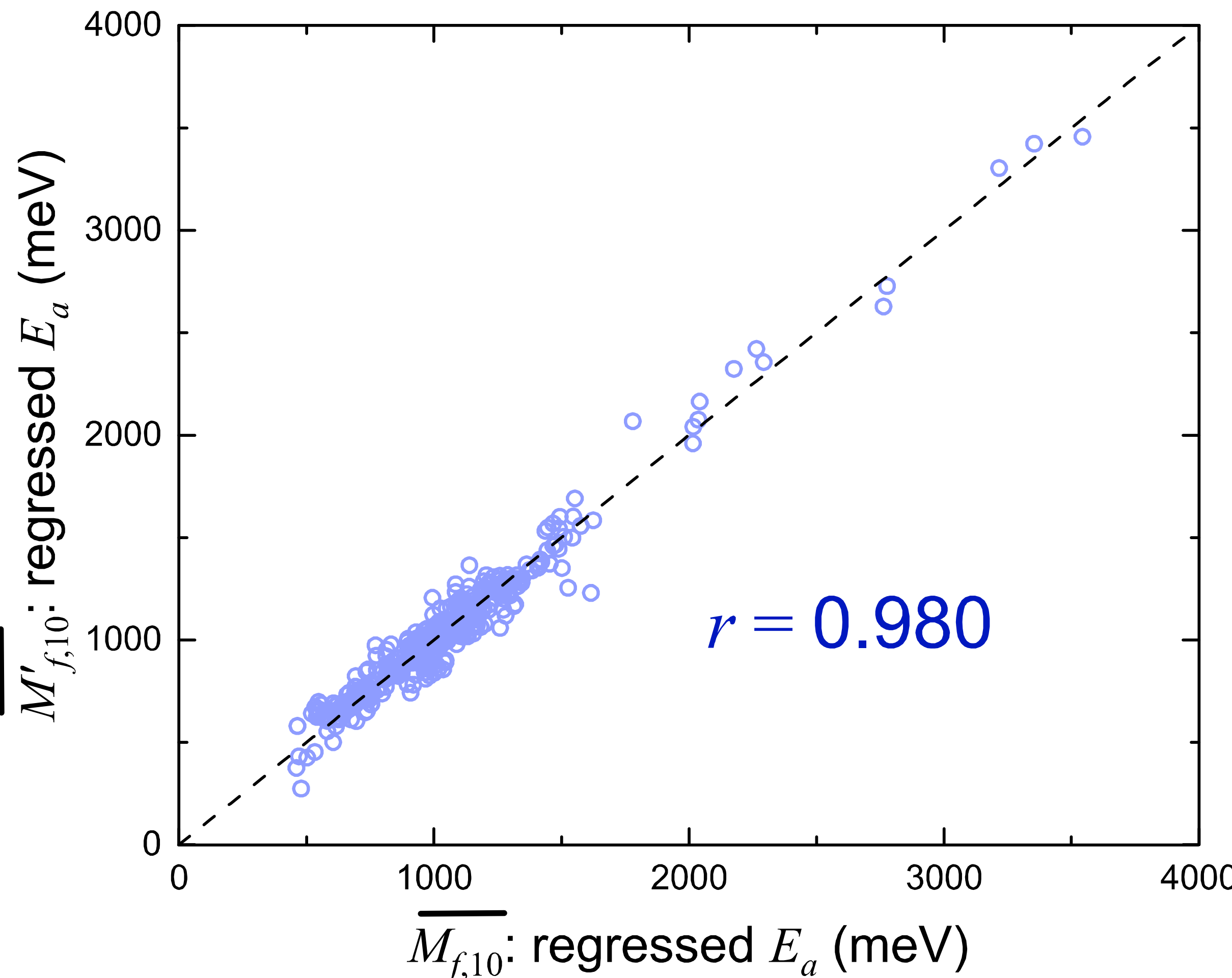

**Supplementary Fig. 8** Correlation between the two independently generated stacking-ensembled models, $\overline{M_{f,10}}$ and $\overline{M_{f,10}}'$, with the Pearson correlation coefficient $r$ shown.

To evaluate the reproducibility of *GoodRegressor*, the stacking-ensembled model for $E_a$was regenerated **from scratch** using new **random** train-validation splits for each individual model. The resulting ensemble, denoted $\overline{M_{f,10}}'$ (dashed), was compared with the original $\overline{M_{f,10}}$. As shown in **Supplementary Fig. 8**, the correlation between the two consensus models is high, with a Pearson correlation coefficient of $r = 0.980$. Beyond internal consistency within the dataset, we also examined external consistency by comparing the predictions of promising candidate materials. Both ensembles, $\overline{M_{f,10}}$ and $\overline{M_{f,10}}'$, identified the identical composition as the most promising: the apatite-type compound $\mathrm{La_{9.5}Si_{5.5}Al_{0.5}O_{26}}$. The predicted $E_a$ were 494 meV (with the $w_{i_f}$-weighted standard deviation across $M_{f,i_f}$, $\sigma_{\forall M}(E_a) = 28.7$ meV; for the detail, see the subsection **"Materials Predictions"** in the **Supplementary Information**) and 491 meV ($\sigma_{\forall M}(E_a) = 39.2$ meV) by $\overline{M_{f,10}}$ and $\overline{M_{f,10}}'$, respectively, with $\log_{10} A = 6.87\ \log_{10}[\mathrm{K \cdot S \cdot cm^{-1}}]$ ($\sigma_{\forall M}(\log_{10} A) = 0.237\ \log_{10}[\mathrm{K \cdot S \cdot cm^{-1}}]$). This composition is a modified analogue of the experimentally reported apatite $\mathrm{Nd_{9.5}Si_{5.5}Al_{0.5}O_{26}}$, which exhibits $E_a = 658$ meV and $\log_{10} A = 7.16\ \log_{10}[\mathrm{K \cdot S \cdot cm^{-1}}]$.[S6] **These analyses demonstrate that, despite the algebraic complexity inherent in symbolic regression models, *GoodRegressor* does not merely overfit noise; rather, it yields stable and meaningful conclusions when provided with the same dataset.**

**Materials Predictions**

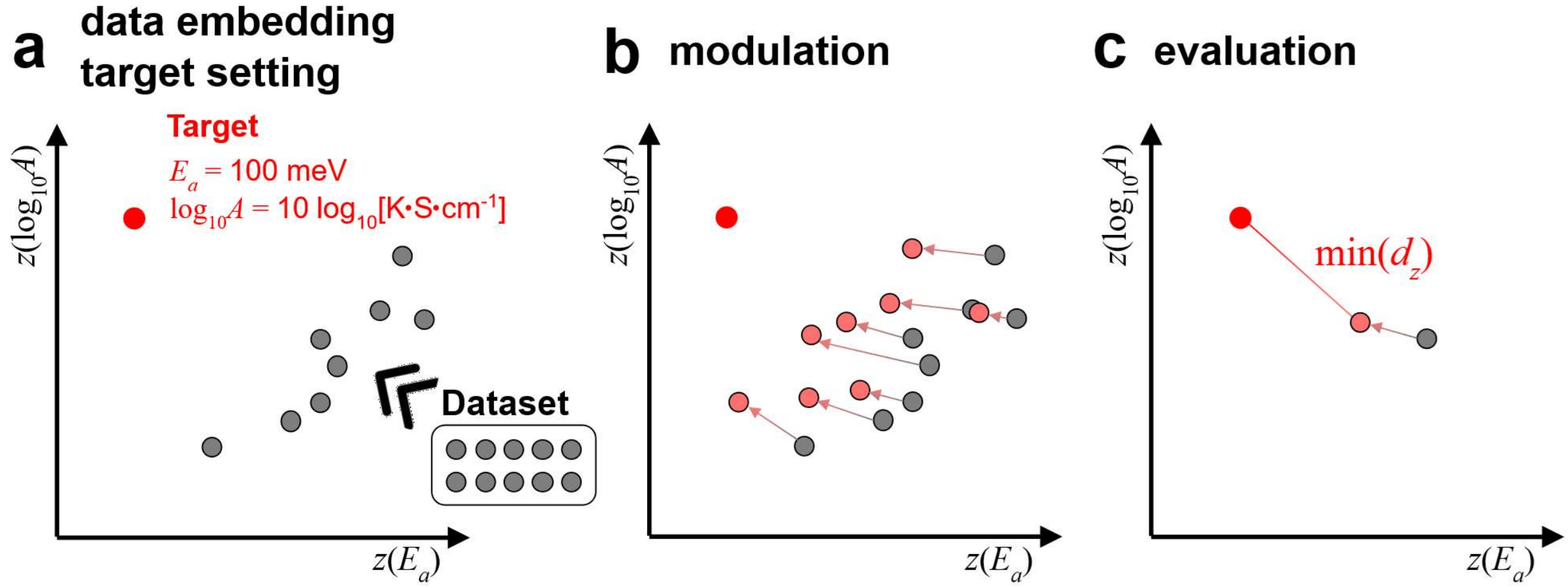

**Supplementary Fig. 9** Schematic of a symbolic regression model-guided materials design workflow. (a) Data embedding and target setting; experimental data (gray points) are mapped into a two-dimensional, $z$-scored target-metric space defined by $E_a$ and $\log_{10} A$. The design target is specified as $\{z(E_{\mathrm{a}} = 100\ \mathrm{meV}), z(\log_{10} A = 10 \log_{10}[\mathrm{K} \cdot \mathrm{S} \cdot \mathrm{cm}^{-1}])\}$ (red points). (b) Modulation; compositions are modified (via atomic/ionic substitutions and content adjustment) to move the gray points as close as possible to the red target points, yielding "modulated" candidates (pink points). (c) Evaluation; the candidate closest to the target is selected by minimizing the $z$-distance $d_z$.

**Supplementary Fig. 9** schematically illustrates the full workflow of the symbolic regression model-guided materials design strategy applied to oxygen-ion conductors. The process aims to identify new compositions that satisfy desired ionic transport properties, characterized by low $E_{\mathrm{a}}$ and high $A$. To ensure both predictive reliability and physical feasibility, the workflow integrates data-driven optimization with physically grounded constraints.

**a. Data embedding and target setting**

483 experimental data points are collected and plotted on a $z$-scored two-dimensional space with the horizontal axis of $z(E_a)$ and the vertical axis of $z(\log_{10} A)$. A target point is defined as $\{z(E_a = 100 \text{ meV}), z(\log_{10} A = 10 \log_{10}[\text{K} \cdot \text{S} \cdot \text{cm}^{-1}])\}$.

**b. Symbolic regression model-guided modulation of composition**

Given the stacking-ensembled models, each chemical composition in the dataset is modulated to explore new design candidates. When a composition is modified by substituting atoms (ions) and their contents, the predicted $E_a$ and $\log_{10} A$ values also change according to the models. Each modulated value is expressed as:

$$\text{(Experimental value)} + \Delta(\text{Predicted value of modified composition} - \text{Predicted value of original composition}) \qquad \text{(S65)}$$

so that the original experimental point acts as an offset for the symbolic regression model-guided perturbation. This modulation part iteratively identifies the atom (ion) substitutions and compositional adjustments that most effectively minimize the distance to the target, where two constraint layers are imposed: structural stability constraint and model reliability constraint.

Imposing the structural stability constraint to prevent instability, the prospective replacement atoms or ions $M'$ for the original species $M$, whether through complete substitution ($M \rightarrow M'$) or 10% doping ($M \rightarrow M_{0.9}M'_{0.1}$), are required to satisfy: (i) same valence as the original element, (ii) electronegativity difference $\Delta < 0.3$, (iii) Shannon ionic radius difference $\Delta < 0.5$ Å.[S2, S3] These constraints may ensure the candidate compositions remain chemically reasonable and structurally feasible.

Under the model reliability constraint, to mitigate the risk of overfitting, the $w_{i_f}$-weighted standard deviations $\sigma_{\forall M}$ of predicted $E_\mathrm{a}$ and $\log_{10} A$ values across multiple regression models are checked. Here, two scenarios were set: $\sigma_{\forall M}(E_\mathrm{a}) < 50$ meV and $\sigma_{\forall M}(\log_{10} A) < 0.5\ \log_{10}[\mathrm{K \cdot S \cdot cm^{-1}}]$. If the variance is large, predictions are considered unreliable and discarded, and if the variance is small within the limits, the candidate is accepted for further evaluation. This internal screening ensures that only stable and reproducible model outputs guide material exploration.

**c. Evaluation of $z$-distance $d_z$**

For each modulated composition that satisfies the above constraints, the $z$-distance to the target is computed as:

$$d_z = \sqrt{w_{E_\mathrm{a}} \cdot (z(E_\mathrm{a}) - z(E_\mathrm{a})_{\mathrm{target}})^2 + w_{\log_{10} A} \cdot (z(\log_{10} A) - z(\log_{10} A)_{\mathrm{target}})^2} \quad \text{(S66)}$$

where the weighting factors are set as: $w_{E_\mathrm{a}} = 0.9$ and $w_{\log_{10} A} = 0.1$. Thus, the model prioritizes minimizing $E_\mathrm{a}$while still considering $\log_{10} A$.

**d. Candidate selection**

All modulated data points are compared based on their computed $d_z$values. The composition(s) with the smallest $d_z$, that is, closest to the target in the $z$-space, are selected as the most promising design candidates. Optionally, multiple top-ranked candidates can be selected for experimental validation or further first-principles screening.

**Promising Material for NASICONs**

**A Zr-free composition, $\mathbf{Na_{3.4}Y_{0.4}Hf_{1.6}Si_2PO_{12}}$, identified as the top-performing candidate by the materials-prediction component of the designer module, is predicted to exhibit $\boldsymbol{\sigma_{\mathrm{Na,RT}} = 3.94 \times 10^{-2}\ \mathrm{S \cdot cm^{-1}}}$** (with the $w_{i_f}$-weighted standard deviation across $M_{f,i_f}$, $\sigma_{\forall M}\left(\log_{10} \sigma_{\mathrm{Na,RT}}\right) = 0.608\ \log_{10}[\mathrm{S \cdot cm^{-1}}]$; for the detail, see the subsection **"Designer Module as a Post-process: Materials Predictions"** in the **Methods** section). This composition is a modified analogue of the experimentally reported $\mathrm{Na_{3.4}Sc_{0.4}Zr_{1.6}Si_2PO_{12}}$, which shows $\sigma_{\mathrm{Na,RT}} = 6.2 \times 10^{-3}\ \mathrm{S \cdot cm^{-1}}$.[S7]

**Promising Materials for Superconducting Oxides**

**A composition, $Cu_{2.8}W_{0.18}Mo_{0.02}Ca_{1.8}Ba_{0.2}Ce_{0.81}Lu_{0.09}Ca_{0.1}O_z$, identified as the top-performing candidate by the materials-prediction component of the designer module, is predicted to exhibit $T_c = 275$ K (with $\sigma_{\forall M}(\log_{10} T_c) = 0.0881 \log_{10}[\mathrm{K}]$).** This composition is a modified analogue of the experimentally reported $Cu_{2.8}Mo_{0.2}Sr_2Y_{0.9}Ca_{0.1}O_z$, which shows $T_c = 71$ K.[S8] Meanwhile, it is noteworthy that the other four top-performing candidates predicted by the model exhibit chemically similar compositions, characterized by the presence of Ca–Cu–(W/Mo)–(Ba)–Ce–(Lu), with predicted critical temperatures in the range of 202 K $< T_c <$ 228 K. Although the exceptionally high value of $T_c = 275$ K for the aforementioned composition is likely attributable to a mathematical artifact of the regression model, the consistency among these chemically related candidates suggests that this compositional space is intrinsically promising and merits further experimental investigation. **Another potential breakthrough, assuming the symbolic regression model is reliable, emerges when allowing substitution (or partial doping) of $Hg^{2+} \rightarrow Ag^{2+}$, a valence state that is relatively rare in naturally occurring compounds, which was not allowed above owing to the unstable nature of $Ag^{2+}$. One representative example is the composition $Ag_{1.314}Hg_{0.146}Br_{1.314}Cl_{0.146}Bi_2Ca_2CaCu_2O_z$ for which the model predicts $T_c = 287$ K (with $\sigma_{\forall M}(\log_{10} T_c) = 0.0848 \log_{10}[\mathrm{K}]$).** This candidate is derived from the experimentally reported compound $Hg_{1.46}Cl_{1.46}Bi_2Sr_2CaCu_2O_z$ with $T_c = 86$ K.[S9] **Notably, the predicted enhancement in $T_c$ upon $Hg^{2+} \rightarrow Ag^{2+}$ substitution appears robust; the model identifies 25 $Ag^{2+}$-containing candidates with predicted $T_c > 200$ K.**

**Symbolic Regression Algorithm: a Structure Loosely Reminiscent of Terence McKenna's *Timewave Zero***

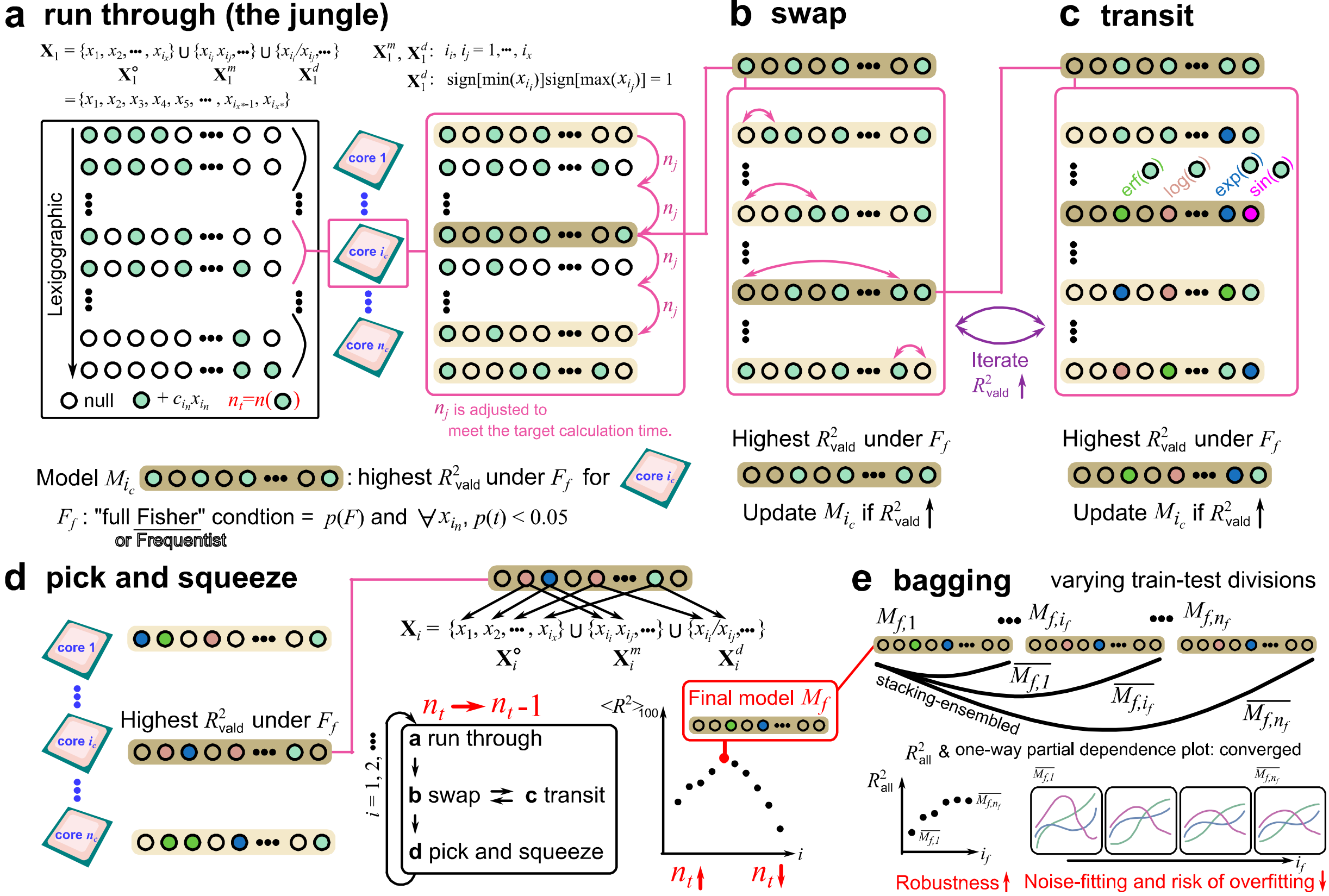

**Supplementary Fig. 10** Schematic workflow of the symbolic regression algorithm implemented in the in-house code, *GoodRegressor*. (a) The "run-through" step explores combinations of $n_t$ descriptor variables and their interactions, distributed across CPU cores in lexicographic order, where the model with the highest $R^2_{\text{vald}}$ value of the validation dataset is selected for each core. (b) The "swap" step replaces less significant variables with inactive ones to improve $R^2_{\text{vald}}$. (c) The "transit" step tests nonlinear transformations to improve $R^2_{\text{vald}}$. (d) Given the fine-tuned model with highest $R^2_{\text{vald}}$ across all the cores, the "pick-and-squeeze" step rebuilds a model with $(n_t \rightarrow) n_t - 1$ variables from an expanded candidate pool including original, interaction,

transformed, and cross-transformed variables, which iterates to maximize $\langle R^2 \rangle_{10}$ with decrease in $n_t$. The “bagging” step repeats the full process with varied data splits to ensemble-average the results, improving robustness and reducing overfitting.

The overall workflow of the regression part is illustrated in **Supplementary Fig. 10**, evoking a structure loosely reminiscent of Terence McKenna’s concept of *timewave zero.* It is noted that the same workflow can also be applied using a beta regression model within this program, which, however, is computationally expensive.[S10-S13]

**Search Code in the Lexicographic Order**

**arrange:** the integer array of which the number of elements is given by the number of taken terms $n_t$.

**jobsize:** the total size of the lexicographic order.

**occup:** the integer array of $\{0, 1\}$. 0 and 1 denote a "taken" and "not taken" term, respectively. For alternative applications, it can be readily expanded to larger integer ranges (e.g., $\{0, 1, 2, \cdots\}$).

**index:** the (target) running number in the lexicographic order.

**Description:** When the jobsize is less than $9 \times 10^{18}$, the **call_XPR** function is invoked. For job sizes in the range $9 \times 10^{18} \leq$ jobsize $\leq 1 \times 10^{4932}$ (or $1 \times 10^{308}$ under MSVC on Windows), the **call_XPR_ld** function is used. The implementations of both functions are provided below.

```
void call_XPR(std::vector<int>* arrange, signed long long int* jobsize, std::vector<int>* occup, signed long long
int* index) {

	signed long long int Ni[2];
	signed long long int Nbar;
	signed long long int index_dynamic = *index;

	for (int iX = 0; iX < (signed)occup->size(); iX++) {
		SigmaMinusXk[iX] = 0;
	}

	for (signed long long int i = 1; i <= arrange->size(); i++) {

		if (i == 1) {
			Ni[0] = *jobsize;
			Ni[1] = (signed long long int) arrange->size();
		}
		else {
			Ni[0] = Nbar;
			Ni[1] = (signed long long int) arrange->size() - i + 1;
		}

		signed long long int Nixk = 0;
		signed long long int Nixk_buf = 0;
		signed long long int DNiXm = 0;
		for (int m = 1; m <= (signed)occup->size(); m++) {
```

```
                        signed long long int multiplier = (signed long long int)(*occup)[m - 1] -
SigmaMinusXk[m - 1];
                        if (multiplier < 0) {
                                multiplier = 0;
                        }
                        DNiXm = Ni[0] * multiplier;
                        if (DNiXm > 0) {
                                DNiXm /= Ni[1];
                        }
                        else {
                                long double DNiXm_ld = (long double)Ni[0] * (long double)multiplier / (long
double)Ni[1];
                                DNiXm = (signed long long int)DNiXm_ld;
                        }
                        Nixk_buf = Nixk;
                        Nixk += DNiXm;
                        if (index_dynamic > Nixk_buf && index_dynamic <= Nixk) {
                                SigmaMinusXk[m - 1]++;
                                (*arrange)[i - 1] = m - 1;
                                index_dynamic -= Nixk_buf;
                                Nbar = DNiXm;
                                break;
                        }
                }

        }

}

bool call_XPR_ld(std::vector<int>* arrange, long double* jobsize, std::vector<int>* occup, long double* index) {

        long double Ni[2];
        long double Nbar;
        long double index_dynamic = *index;

        for (int iX = 0; iX < (signed)occup->size(); iX++) {
                SigmaMinusXk_ld[iX] = 0;
        }

        bool stable = true;

        for (signed long long int i = 1; i <= arrange->size(); i++) {

                if (i == 1) {
                        Ni[0] = *jobsize;
                        Ni[1] = (signed long long int) arrange->size();
                }
                else {
                        Ni[0] = Nbar;
                        Ni[1] = (signed long long int) arrange->size() - i + 1;
                }

                long double Nixk = 0;
                long double Nixk_buf = 0;
                long double DNiXm = 0;
                bool m_taken = false;
```

```
		for (int m = 1; m <= (signed)occup->size(); m++) {
			signed long long int multiplier = (signed long long int)(*occup)[m - 1] -
SigmaMinusXk_ld[m - 1];
			if (multiplier < 0) {
				multiplier = 0;
			}
			DNiXm = Ni[0] * multiplier;
			if (DNiXm > 0) {
				DNiXm /= Ni[1];
			}
			else {
				long double DNiXm_ld = (long double)Ni[0] * (long double)multiplier / (long
double)Ni[1];
				DNiXm = (signed long long int)DNiXm_ld;
			}
			Nixk_buf = Nixk;
			Nixk += DNiXm;
			if (index_dynamic > Nixk_buf && index_dynamic <= Nixk) {
				SigmaMinusXk_ld[m - 1]++;
				(*arrange)[i - 1] = m - 1;
				index_dynamic -= Nixk_buf;
				Nbar = DNiXm;
				m_taken = true;
				break;
			}
		}
		if (!m_taken) {
			int unoccup_count = 0;
			for (int iarr = 0; iarr < (signed)arrange->size(); iarr++) {
				if (unoccup_count < (*occup)[0]) {
					(*arrange)[iarr] = 0;
					unoccup_count++;
				}
				else {
					(*arrange)[iarr] = 1;
				}
			}
			break;
			stable = false;
		}

	}

	return stable;

}
```

**Supplementary References**